\documentclass[journal]{IEEEtran}
\usepackage{subfigure}
\usepackage{cite}

\ifCLASSINFOpdf
   \usepackage[pdftex]{graphicx}

\else

\fi

\usepackage[cmex10]{amsmath}
\usepackage{url}

\usepackage{amssymb}

\usepackage{colortbl}

\usepackage{epstopdf}

\usepackage{relsize}

\begin{document}

\title{Joint Scheduling and Transmission Power Control in Wireless Ad Hoc Networks}
%Coordination-based Medium Access Control for Wireless Ad Hoc Networks
\author{Kamal~Rahimi Malekshan,~\IEEEmembership{Member,~IEEE,}
        and~Weihua~Zhuang,~\IEEEmembership{Fellow,~IEEE}% <-this % stops a space
        \vspace{-.45cm}
\thanks{This work was supported by a research grant from the Natural Sciences and Engineering Research Council (NSERC) of Canada.

K. Rahimi Malekshan is with the Research and Development Department of Siemens, Toronto, Canda (e-mail: kamal.malekshan@siemens.com).

W. Zhuang is with the Department of Electrical and Computer Engineering, University of Waterloo, Canada (e-mail: wzhuang@uwaterloo.ca).
%
%Digital Object Identiﬁer ...
}
}
\maketitle

\begin{abstract}
%\boldmath
In this paper, we study how to determine concurrent transmissions and the transmission power level of each link to maximize spectrum efficiency and minimize energy consumption in a wireless ad hoc network. The optimal joint transmission packet scheduling and power control strategy are determined when the node density goes to infinity and the network area is unbounded. Based on the asymptotic analysis, we determine the fundamental capacity limits of a wireless network, subject to an energy consumption constraint. We propose a scheduling and transmission power control mechanism to approach the optimal solution to maximize spectrum and energy efficiencies in a practical network. The distributed implementation of the proposed scheduling and transmission power control scheme is presented based on our MAC framework proposed in \cite{Kamal3}. Simulation results demonstrate that the proposed scheme achieves 40\% higher throughput than existing schemes. Also, the energy consumption using the proposed scheme is about 20\% of the energy consumed using existing power saving MAC protocols.
\end{abstract}

\begin{IEEEkeywords}
 Ad hoc networks, radio spectrum efficiency, energy efficiency, spatial reuse, transmission power control, medium access control (MAC).
\end{IEEEkeywords}

\IEEEpeerreviewmaketitle

\section{Introduction}
The increasing number of mobile devices and volume of mobile Internet traffic necessitate dense deployment of Internet access points (APs) in an ad hoc manner to increase network capacity via shorter communication links \cite{Dense-SmallCell}. Also, diverse peer-to-peer communications \cite{Device-to-Device-Survey,Machin-to-Machin-Survey} are emerging to increase spectrum and energy efficiencies via shorter communication links and to interconnect several billion physical objects and integrate them into the existing networks. Such a dense and dynamic network of mobile nodes and APs and diverse peer-to-peer communications require to establish an effective ad hoc network to efficiently utilize radio spectrum and to minimize energy consumption.

In a wireless network, the data rate and energy consumption of a link depend on the transmission power level of the source node, the distance between source and destination, and the amount of interference at the destination node. The amount of interference at a destination depends on the distance to interfering source nodes and their transmission power level. Thus, the achievable data rate and energy consumption of transmitting links are interrelated. The set of concurrent transmissions and the transmission power level of each source should be properly determined to efficiently utilize radio spectrum and reduce energy consumption. In addition, a radio interface consumes a significant amount of energy in the idle mode in which it is not transmitting/receving a packet. The energy consumption can be reduced by putting the radio interface in a sleep mode, however, the awake and active times of the radio interface should be properly scheduled to avoid missing incoming packets \cite{Kamal3, DPSM, Kamal2, TMMAC,Kamal}.

Although increasing spatial reuse allows more concurrent transmissions, it also decreases the signal-to-noise-plus interference-ratio (SINR) at the receivers. Therefore, the data rate of each transmission decreases as a result of a lower SINR. The trade-off between the increased spatial reuse and the decreased data rate has been studied in \cite{Kim,Kim2} when using a CSMA/CA MAC. It is shown that the network capacity depends only on the ratio of the transmission power level to the carrier sensing threshold (i.e., carrier sensing range). It is proposed that all nodes use a same carrier sensing threshold and each source node adjusts its transmission power level based on its distance from the destination. However, when only carrier sensing is used, the transmission rates must be adjusted for the worst case interference to ensure successful reception of packets at the receiver. As a result, the transmission power control schemes (in which nodes independently choose their transmission power levels) cannot fully utilize the network capacity. Also, the CSMA based MAC protocols provide poor spatial spectrum reuse due to the \begin{it}hidden\end{it} and \begin{it}exposed\end{it} node problems \cite{Kamal3,HNEN-Jiang}. Centralized scheduling and transmission power control for wireless ad hoc networks are proposed in \cite{STPC-C,STPC-LS}.

The optimal scheduling and transmission power control to maximize total throughput in a two-cell two-link wireless network have been studied in \cite{TWO-Cell}. In the network with two links, maximizing total throughput leads to binary power control. That is, each link should transmit at either the maximum power level or the minimum power level \cite{TWO-Cell}. Motivated by the optimality of binary power control, the binary power control is also proposed for multi-cell networks with more than two links in \cite{Binary-Power-Control}.

The effect of transmission power level on total energy consumption depends on the energy consumption pattern of the radio interface \cite{Interference-Cellular, PointToPoint, Principles, COMPOW}. The energy consumption has two components: the energy consumed in the radio interface circuit, and the energy consumed in the amplifier. When the energy consumption in the amplifier dominates the energy consumed at the radio interface circuit, the energy consumption per transmitted data bit in a two-link network can be reduced by decreasing the transmission power level \cite{Interference-Cellular}. However, when the energy consumption in the radio interface circuit is much larger than the energy consumption in the amplifier, minimizing the energy consumption per transmitted data bit in a two-link network is equivalent to maximizing network throughput \cite{Interference-Cellular}. Generally, the transmission power level for minimal energy consumption depends on the energy consumption pattern of the radio interface and the network condition. Thus, transmission at the minimum power level (as in \cite{PCM, PCMA, Intelligent-medium, PCDC}) does not always reduce the energy consumption.

In \cite{Kamal3}, we present a novel MAC scheme for a wireless ad hoc network. All node transmissions are dynamically scheduled by a set of coordinator nodes that are distributed over the network coverage area. A coordinator node monitors source nodes' transmission requests in its proximity, actively exchanges scheduling information with its adjacent coordinators, and periodically determines contention-free transmission/reception times for nodes in its vicinity. For each scheduled transmission a proper space area around the receiver node is reserved to guarantee the required link SINR and enhance spatial spectrum reuse. Moreover, the deterministic data transmission time allows nodes to stay awake only when they are transmitting/receiving a packet to minimize idle-listening energy consumption.

In this paper, we study efficient joint transmission scheduling and power control in a wireless ad hoc network. We show that the asymptotic optimal scheduling and transmission power control can be determined when node density in the network goes to infinity and the network area is unbounded. By analyzing the asymptotic optimal solution, we determine the fundamental limits of maximum spectrum and energy efficiencies in a wireless network. To approach the maximum spectrum and energy efficiencies in a practical network, we assign a transmission power level and a target interference power level to each link, which are determined based on the asymptotic optimal values. The concurrent transmissions at each time slot are scheduled such that the actual power of interference at the scheduled destination nodes are close to the target interference levels for efficient spectrum and energy utilization. We present a distributed implementation of the proposed scheduling and transmission power control scheme based on our MAC framework proposed in \cite{Kamal3}.

The main contributions of this paper can be summarized as follows:

1) We analyze asymptotic joint optimal scheduling and transmission power control, and determine the fundamental limits of network capacity, subject to an energy efficiency constraint;

2) Based on the asymptotic optimal solution, we propose a novel scheduling and transmission power control framework to approach maximum spectrum and energy efficiencies in a practical network. Also, we present a distributed implementation of the proposed scheme using only local network information;

3) The throughput and energy consumption of our proposed scheduling and transmission power control framework are evaluated in comparison with existing schemes. A new scheduling efficiency metric is introduced to compare the efficiency of different schemes with the asymptotic optimal solution.

The rest of this paper is organized as follows: The system model is presented in Section \ref{SysMod}. In Section \ref{AsymOpt}, we analyze asymptotic joint optimal scheduling and transmission power control and determine the maximum spectrum and energy efficiencies in the wireless network. We propose a scheduling and transmission power control framework to approach the optimal solution in a practical network in Section \ref{JSTPC}. Simulation results are presented in Section \ref{SimRes}. Finally, Section \ref{Summary-P3} concludes this paper.

\section{System Model}\label{SysMod}
Consider a wireless ad hoc network where all network nodes use a shared radio channel for transmissions. We focus on single-hop transmissions as, at the MAC layer, each node communicates with one or more of its one-hop neighboring nodes\footnote{Note that the end-to-end communication link between two nodes may compose of one or several hops whose path/relays can be decided in the network layer (using a routing algorithm). That is, a link at the MAC layer can be a single-hop of an end-to-end multi-hop transmission path.}. Nodes are randomly distributed in the network area and the destination of each source node is randomly selected from the rest nodes within maximum data transmission distance $d_{\max}$. Let $L$ denote the number of links and $l\in\{1,2,...,L\}$ denote a link; The source and destination nodes of link $l$ are denoted by $S_l$ and $D_l$, respectively. Network links are considered to be directional (i.e., transmission from a source to a destination node). Bidirectional communications (such as a TCP link) between two nodes are handled by scheduling two different directional links. We denote the distance from the source node of link $l$ to the destination node of link $k$ by $d_{lk}$, and the associated channel gain is $h_{lk}=c{d_{lk}}^{-\alpha}$, where $c$ is a constant and $\alpha$ is the path-loss exponent\footnote{We assume that physical-layer channel coding deals with channel fading.}.

Time is partitioned into slots of constant durations. Consider a scheduling interval of $T$ slots, and let $t\in{\{1, 2, ...,T\}}$ denote time slot index\footnote{The scheduling interval should be determined based on data traffic and network dynamics. A very large scheduling interval causes slow adaptation to data traffic and network changes, while a small scheduling interval leads to higher scheduling overhead due to more frequent scheduling/signaling slots.}. We assume that $d_{lk}$, with $l,\, k \in\{1,2,...,L\}$, is constant over $T$ time slots. Let $\bar{\boldsymbol{\gamma}}={[\gamma_{lt}]}_{L\times T}$ denote the transmission power matrix, where $\gamma_{lt}$ denotes the transmission power level of source node of link $l$ at time slot $t$. Let $\bar{\boldsymbol{u}}={[u_{lt}]}_{L\times T}$ denote the scheduling matrix, where $u_{lt}=1$ if link $l$ is scheduled for transmission at time slot $t$ and $u_{lt}=0$ otherwise. A scheduled link transmits a data packet during a time slot that is scheduled. The SINR at the destination of link $l$ at slot $t$ is given by $\eta_{lt}=\frac{u_{lt} \gamma_{lt}h_{ll}}{N_0+\sum_{k\neq l}u_{kt} \gamma_{kt}h_{kl}}$, where $N_0$ is background noise power and $\sum_{k\neq l}u_{kt}\gamma_{kt}h_{kl}\triangleq I_{lt}$ is the power of interference at the destination. The achievable channel rate in bit/s/Hz over link $l$ at slot $t$, using \begin{it}Shannon formula\end{it}, is $R_{lt}=\log_2(1+\eta_{lt})$ and the average data rate at link $l$ can be written as $R_{l}=\frac{1}{T}\sum_{t=1}^{T}R_{lt}$.

A radio interface can be in transmit, receive, idle and sleep modes. The power consumption of a radio interface in the transmit mode to transmit at power level $\gamma$ is $\Gamma_c+g_a\gamma$, where $\Gamma_c$ is the circuit power consumption and $g_a>1$ is the inverse of the power efficiency of radio interface amplifier. The power consumption in the receive and idle modes is $\Gamma_c$ and in the sleep mode is $\Gamma_0$. Each node puts its radio interface in sleep mode when it is not transmitting/receiving data to save energy. Thus, the sum of power consumption (in Joule/s) at the source and destination nodes of link $l$ at slot $t$ is $P_{lt}=u_{lt}\times(2\Gamma_c+g_a\gamma_{lt})+(1-u_{lt})\times(2\Gamma_0)$ and the average power consumption at link $l$ is $P_{l}=\frac{1}{T}\sum_{t=1}^{T}P_{lt}$. The average energy consumed per transmitted bit (in Joule/(bit/Hz)) at link $l$ can be written as $E_{l}=P_l/R_l$.

Joint optimal scheduling and transmission power control are to find a scheduling matrix and a transmission power matrix that maximize the network objective function, given by
\begin{equation}\label{C4-0e}
  \begin{array}{l}
  \max\limits_{\bar{\boldsymbol{u}},\bar{\boldsymbol{\gamma}}}\;{\sum_{l=1}^{L}{w_l R_l }}\\
              \quad \text{s. t.  }\, R_l\leq \hat{R}_l,\; E_l\leq \hat{E}_l,\; \scalebox{.8}{$l\in\{1,2,...,L\}$}\\
  \end{array}
\end{equation}
where $w_l\in [0,\infty)$ is the weighting factor of data rate of link $l$, $\hat{R}_l$ denotes the maximum required data rate at link $l$, and $\hat{E}_l$ denotes the maximum energy consumption per bit constraint at link $l$. To find an optimal solution in (\ref{C4-0e}), we need to solve a non-convex mixed integer non-linear problem, which is known to be NP-hard \cite{RAND, NP-hard}.

\section{Asymptotic Joint Optimal Scheduling and Transmission Power Control}\label{AsymOpt}
\begin{figure}
  \centering
  \includegraphics[width=3.2in, trim=3.8cm .5cm 1.5cm 1.5cm, clip=true]{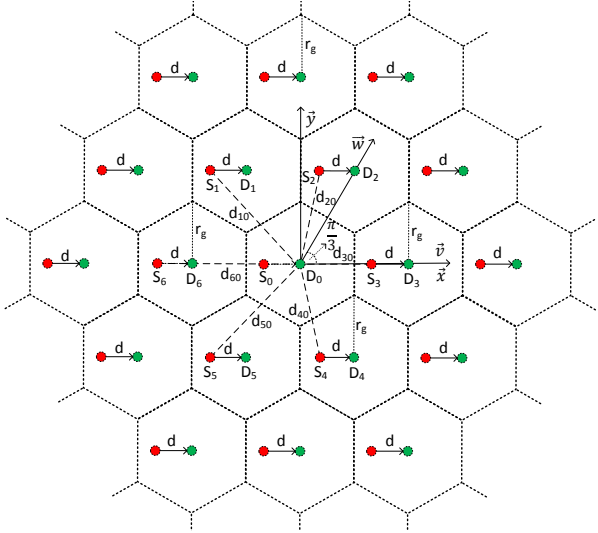}
  \caption{Illustration of symmetric scheduling}\label{SymNetwZ3}
\end{figure}

In this section, we study scheduling and transmission power control in the wireless network as the node density goes to infinity and the network area is unbounded. Consider a symmetric link scheduling in an unbounded network area as illustrated in Figure \ref{SymNetwZ3}. The network area is partitioned into equal size hexagonal cells and a link is scheduled inside each cell. The source and destination distance is the same for all links and the position of every scheduled link with respect to all other scheduled links is identical. Due to the symmetry of scheduled links, the optimal transmission power should be the same for every scheduled link. Thus, the asymptotic optimal joint scheduling and transmission power control are to find a cell size and a transmission power level that maximize the network objective function. In the following, we analyze the spectrum and energy efficiencies in the network as the cell size and transmission power level vary, in order to determine the optimal scheduling and transmission power control.

\begin{figure}
  \centering
  \includegraphics[width=3.2in, trim=.25cm 1.25cm .2cm 2cm, clip=true]{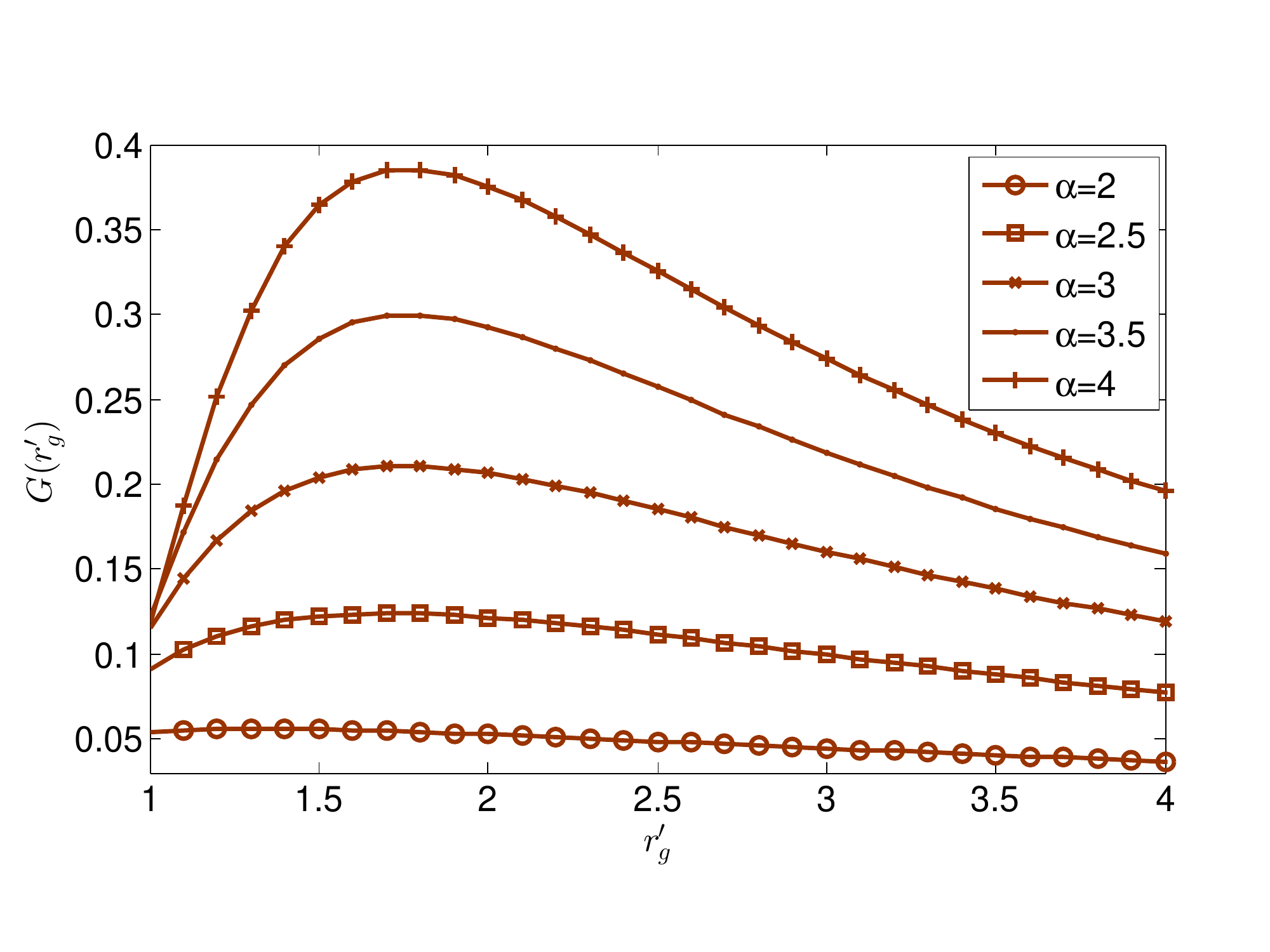}
  \caption{Plot of function $G(\cdot)$ for different path-loss exponent values}\label{PlotGh}
\end{figure}

Let $d$ denote the distance between the source and destination of a link, $r_g$ the distance between the center and a vertex of a cell, and $\gamma$ the transmission power of every scheduled source node. The signal power at a destination node is $\gamma^{(r)}=c \gamma {d}^{-\alpha}$. Let $d_{i0}, i\in{\{1,2,...\}}$, denote the distance from the source node of an interfering link to the destination node of a target link. Using unity vectors $\vec{v}$ and $\vec{w}$, we have
\begin{equation}\label{C4-2b}
  d_{i0}=\big|\big|(m\sqrt{3}r_g-d)\vec{v}+n\sqrt{3}r_g\vec{w}\big|\big|
\end{equation}
for some $(m,n)\in{\{...,-2,-1,0,1,2,...\}}^2, (m,n)\neq(0,0)$, where $||\cdot||$ denotes the Euclidian distance. By changing coordinates in (\ref{C4-2b}), we have
\begin{multline}\label{C4-2c}
  d_{i0}=\big|\big|(m\sqrt{3}r_g+n\sqrt{3}r_g/2-d)\vec{x}+(3nr_g/2)\vec{y}\big|\big|
  \\=\Big[\big(m\sqrt{3}r_g+n\sqrt{3}r_g/2-d\big)^2+\big(3nr_g/2\big)^2\Big]^{1/2}.
\end{multline}
The interference power at a destination node can be calculated as $I=\sum_{i=1}^{\infty} c\gamma  {d_{i0}}^{-\alpha}$. With the assumption that $I \gg N_0$, the SINR at a destination node can be calculated as
\begin{multline}\label{C4-3}
  \eta=\frac{\gamma^{(r)}}{N_0+I}\approx\frac{\gamma^{(r)}}{I}=\frac{c\gamma{d}^{-\alpha}}{\sum_{i=1}^{\infty} c\gamma  {d_{i0}}^{-\alpha}}\\=\frac{1}{\sum_{(m,n)\neq(0,0)}{{\Big[\big(\frac{m\sqrt{3}r_g}{d}+\frac{n\sqrt{3}r_g}{2d}-1\big)^2+\big(\frac{3nr_g}{2d}\big)^2\Big]}^{-\alpha/2}}} \\\triangleq F(\tfrac{r_g}{d}).
\end{multline}
Also, with frequency reuse, the network space occupied by each scheduled link is given by
\begin{equation}\label{C4-4}
  S=3\sqrt{3}/2\times r_g^2.
\end{equation}
Using (\ref{C4-3}) and (\ref{C4-4}), the total data rate (bit/s/Hz) per unit network area can be written as
\begin{equation}\label{C4-5}
  \tilde{R}=\frac{\log_2(1+\eta)}{S}=\frac{1}{d^2}\times \frac{\log_2\left(1+F(r_g/d)\right)}{3\sqrt{3}/2\times\big(r_g/d\big)^2}.
\end{equation}
According to (\ref{C4-5}), the total data rate depends on the ratio $r_g/d\triangleq r'_g$, and can be maximized by choosing $r'_g$ to maximize $\frac{\log_2\left(1+F(r'_g)\right)}{3\sqrt{3}/2\times{r'_g}^2}\triangleq G(r'_g)$. Function $G(\cdot)$ is plotted in Figure \ref{PlotGh} for different path-loss exponent values. Also, the maximum achievable data rate is inversely proportional to the square of the link distance, $1/d^2$. On the other hand, energy consumption per transmitted data bit (Joule/(bit/Hz)) is
\begin{equation}\label{C4-6}
  E=\frac{ (2\Gamma_c+g_a \gamma)/S}{\tilde{R}}=\frac{2\Gamma_c+g_a \gamma}{\log_2\left(1+F(r_g/d)\right)}
\end{equation}
where $2\Gamma_c+g_a \gamma$ denotes the sum of power consumption in the source and destination nodes of a scheduled link (with the assumption that power consumption in a source node is $\Gamma_c+g_a \gamma$, in a destination node is $\Gamma_c$ and in a non-scheduled node is negligible. According to (\ref{C4-6}), the energy consumption per transmitted data bit decreases as the distance between scheduled links increases (i.e., as $r_g$ increases).

We set the objective of joint scheduling and transmission power control to maximize the total data rate per unit network area, while keeping the amount of consumed energy per transmitted data bit below a threshold, $\hat{E}$, as an energy efficiency constraint. That is,
\begin{equation}\label{C4-8}
  \begin{aligned}
  {\max\limits_{\gamma,r_g}\; \frac{1}{d^2}\times \frac{\log_2\left(1+F(\frac{r_g}{d})\right)}{\frac{3\sqrt{3}}{2}\times\left(\frac{r_g}{d}\right)^2}} \qquad \qquad \\[3pt]
                  \quad\text{s. t.       }\; \displaystyle\frac{2\Gamma_c+g_a \gamma}{\log_2\left(1+F(\frac{r_g}{d})\right)}\leq \hat{E}, \;F(\frac{r_g}{d}) \geq \eta_{\min}
  \end{aligned}
\end{equation}
where $\eta_{\min}$ is the minimum required SINR at a destination node for successful signal detection. The objective function in (\ref{C4-8}) is consistent with (\ref{C4-0e}) in which $w_l=1$, $\hat{R}_l=\infty$, and $\hat{E}_{l}=\hat{E}$, for every link $l$. We numerically solve (\ref{C4-8}) using a brute-force search over discrete values of $\gamma$ and $r_g$. Also, an alternative way to solve (\ref{C4-8}) based on the Lagrangian multipliers method and Karush-Kuhn-Tucker (KKT) conditions is discussed in the Appendix. Figure \ref{RvsE} shows spectrum efficiency and energy consumption per bit with optimized transmission power and cell size, as the the energy consumption constraint $\hat{E}$ varies.

\begin{figure}
  \centering
  \includegraphics[width=3.2in, trim=.25cm 1.25cm .2cm 2cm, clip=true]{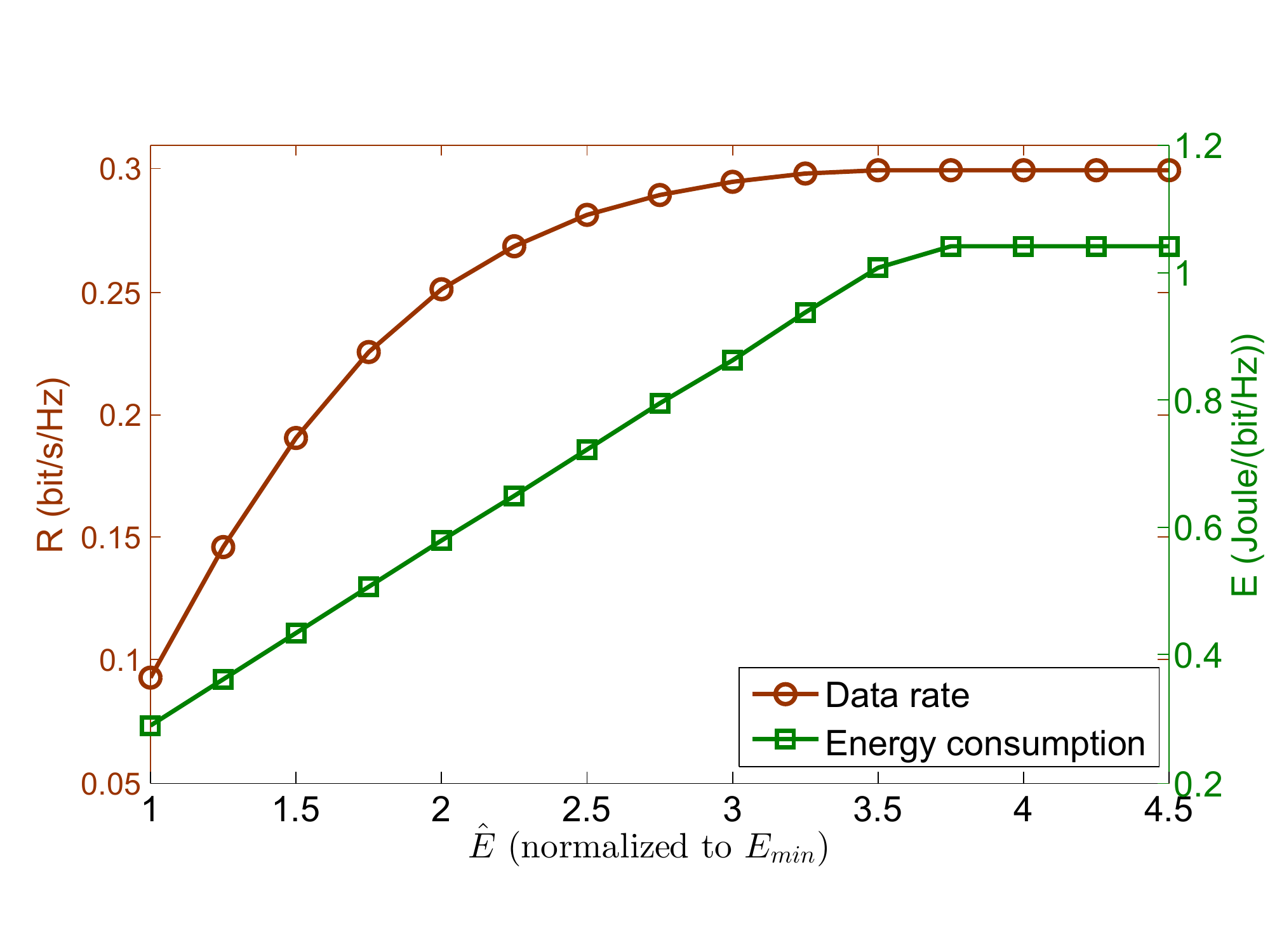}
  \caption{Data rate per unit of network area versus energy consumption per transmitted bit as $\hat{E}$ in the network objective function varies ($\Gamma_c=1.25$ mW, $\gamma \in[1,100]$ mW, $g_a=10$, $\alpha=3.5$, $d=1$, $r_g\in[d,4d]$.)}\label{RvsE}
\end{figure}

\section{Scheduling and Transmission Power Control}\label{JSTPC}
In a practical wireless network, scheduled links likely can not be placed in a symmetric manner, because the node density is finite and the link distances are not identical. Also, scheduling and transmission power control should be adaptive, as node location and traffic load vary over time. As discussed in Section \ref{SysMod}, the optimal scheduling and transmission power control are in general solutions of an NP-hard problem. Thus, we develop a heuristic scheduling and transmission power control framework based on the asymptotic optimal solution.

The data rate and energy consumption of a link depend on the transmission power of the source and the power of interference at the destination node. We schedule links for transmissions such that the transmission power of source nodes and the power of interference at destination nodes follow the asymptotic optimal values. For this purpose, we assign a transmission power level to the source and a target interference power level to the destination of each link, which follow the values that maximize asymptotic spectrum efficiency while satisfying the energy consumption per bit constraint of the link. Then, we schedule concurrent links for transmissions such that the actual power of interference at the destination of each scheduled link is not larger than but as close as possible to the determined target interference power of the link. If the actual interference at a destination node is more than the target interference power, the data will not be successfully decoded at receiver (because the actual SINR at the destination node will be lower than the targeted SINR value used to adjust transmission data rate at the source node). However, it is desired to schedule links such that the actual interference at destinations are close to the target interference of the schedule links in order to allow more concurrent transmissions.

\subsection{Transmission power and target interference power} \label{TPIBC}
We determine the transmission power and target interference power for a link based on the levels that maximize the asymptotic spectrum efficiency (data rate per unit area) while maintaining the energy consumption per bit of the link below a threshold. Using (\ref{C4-3}), we have
\begin{equation}\label{C4-5fa}
  \frac{r_g}{d}=F^{-1}(\frac{\gamma}{I}).
\end{equation}
By substituting (\ref{C4-5fa}) in (\ref{C4-5}) and (\ref{C4-6}), for transmission between a pair of source and destination nodes with distance $d_{ll}$, setting the transmission power to $\gamma_l$ and the target interference power to $\tilde{I}_l$ provides the asymptotic spectrum efficiency
\begin{equation}\label{C4-5f}
  \tilde{R}_l=\frac{1}{{d_{ll}}^2}\times \frac{\log_2\big(1+\frac{c \gamma_l {d_{ll}}^{-\alpha}}{{\tilde{I}_l}}\big)}{\frac{3\sqrt{3}}{2}\times\Big[{F}^{-1}\big(\frac{c \gamma_l {d_{ll}}^{-\alpha}}{{\tilde{I}_l}}\big)\Big]^2}
\end{equation}
and energy consumption per transmitted bit
\begin{equation}\label{C4-6f}
  E_l=\displaystyle\frac{2\Gamma_c+g_a \gamma}{\log_2\big(1+\frac{c \gamma_l {d_{ll}}^{-\alpha}}{{\tilde{I}_l}}\big)}.
\end{equation}
According to (\ref{C4-5f}), the asymptotic spectrum efficiency is inversely proportional to the link distance square, ${d_{ll}}^2$, and depends on the ratio of transmission power to target interference power, $\gamma_l/\tilde{I}_l$. Also, the optimal ratio $\gamma_l/\tilde{I}_l$ depends on the link distance, $d_{ll}$.

\begin{figure}
  \centering
  \includegraphics[width=3.0in, trim=6cm 9.8cm 9.4cm 7cm, clip=true]{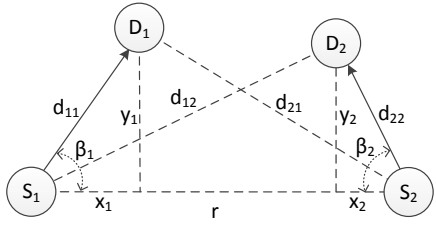}
  \caption{A two-link network}\label{TwoLinkFig}
\end{figure}

In a practical wireless network, the distances between the source and destination nodes of different links are different in general. Thus, the desired ratios of transmission power to interference power for links with different distances are different. Given the different desired ratios of transmission power to interference power for the links, the transmission power and target interference power values should be prudently chosen such that links can be scheduled with actual interference power close to the target interference level at every scheduled link for efficient spatial spectrum reuse. The transmission power of a link determines the minimum distance between its source node and the destination of rest scheduled links, however, its target interference level determines the minimum distance between its destination node and the source node of rest scheduled links. To illustrate, consider a two-link network depicted in Figure \ref{TwoLinkFig}. As transmission power of $S_1$ increases, the amount of interference imposed by $S_1$ on $D_2$ also increases. Thus, to maintain a target interference level, $d_{12}$ must be increased. Similarly, decreasing the target interference in $D_1$ requires larger distance $d_{21}$ to reduce the imposed interference from $S_2$ on $D_1$. Therefore, it is desired that a link with higher transmission power to also have lower target interference level for efficient spatial spectrum reuse. To study how to choose the transmission power and target interference value of different links, we consider a two-link network as illustrated in Figure \ref{TwoLinkFig}. We assume that $\beta_1$ and $\beta_2$ are independent and uniformly distributed in $[0,2\pi]$. We also assume that the distances between the source and destination of the links, $d_{11}$ and $d_{22}$, in different two-link network realizations, are independent and have an identical distribution. Let $E(d_{11})=E(d_{22})=m_1$ and $E({d_{11}}^2)=E({d_{22}}^2)=m_2$. We consider the distance between the two source nodes ($r$ in Figure \ref{TwoLinkFig}) as a measure of the space occupied by the two scheduled links. Thus, it is desired to minimize the expected distance $r$ (over random realization of $\beta_1$, $d_{11}$, $\beta_2$ and $d_{22}$) to minimize the average occupied space for the scheduled links and, as a result, maximize spatial spectrum reuse. Both links can be scheduled concurrently only if the actual interference power at each link is not greater than its target level. That is
\begin{equation}\label{C4-13}
 I_j=c\gamma_i {d_{ij}}^{-\alpha} \leq \tilde{I}_j \Rightarrow d_{ij} \geq \Big({\frac{c\gamma_i}{\tilde{I}_j}}\Big)^{\frac{1}{\alpha}}, \scalebox{.8}{$(i,j)\in\big\{(1,2),(2,1)\big\}$}
\end{equation}
According to Figure \ref{TwoLinkFig}, we have
\begin{equation}\label{C4-13b}
d_{ij}=\sqrt{\left(r-x_j\right)^2+{y_j}^2}=\sqrt{r^2-2rd_{jj}\cos(\beta_j)+{d_{jj}}^2}.
\end{equation}
By substituting (\ref{C4-13b}) in (\ref{C4-13}), the required conditions to schedule both links concurrently can be written as
\begin{equation}\label{C4-13c}
 r^2-2rd_{jj}\cos(\beta_j)+{d_{jj}}^2\geq \Big({\frac{c\gamma_i}{\tilde{I}_j}}\Big)^{\frac{2}{\alpha}}, \scalebox{.8}{$(i,j)\in\big\{(1,2),(2,1)\big\}$}.
\end{equation}
Taking expectation (with respect to $\beta_j$ and $d_{jj},$ \scalebox{.8}{$j\in\{1,2\}$}) from both sides of (\ref{C4-13c}), we obtain
\begin{equation}\label{C4-15}
  E(r^2)\geq \max{\Big[\Big({\frac{c\gamma_1}{\tilde{I}_2}}\Big)^{\frac{2}{\alpha}}-m_2,\Big({\frac{c\gamma_2}{\tilde{I}_1}}\Big)^{\frac{2}{{\alpha}}}-m_2\Big]}.
\end{equation}
According to (\ref{C4-15}), the expected square of distance, $E(r^2)$, increases as the transmission power levels increase and target interference power levels decrease. Also, $E(r^2)$ can be decreased by setting
\begin{equation}\label{C4-16}
  \Big({\frac{c\gamma_1}{\tilde{I}_2}}\Big)^{\frac{2}{\alpha}}-m_2=\Big({\frac{c\gamma_2}{\tilde{I}_1}}\Big)^{\frac{2}{\alpha}}-m_2 \Rightarrow \gamma_1\times \tilde{I}_1 = \gamma_2\times \tilde{I}_2.
\end{equation}
Thus, the average occupied space for scheduling links is decreased (i.e., actual interference power levels are close to the target interference power levels in both links) when the product of transmission power and target interference power is identical for every link. This constraint ensures that a link with greater transmission power to target interference level ratio is optimally assigned both a higher transmission power and a lower target interference for efficient spatial spectrum reuse. Motivated by the analysis for the two-link network, we maintain the product of transmission power and target interference power at a fixed value for all links in the network.

Therefore, we determine the transmission power $\gamma^*_l$ and target interference power ${\tilde{I}}_l^*$ for link $l$, such that asymptotic spectrum efficiency (\ref{C4-5f}) is maximized subject to energy consumption per bit (\ref{C4-6f}) smaller than a threshold, while maintaining the product of transmission power and target interference power at a fixed value.  Thus, transmission power $\gamma^*_l$ and target interference power ${\tilde{I}}_l^*$ are calculated as follows.
\begin{equation}\label{C4-18b}
  \begin{aligned}
  {[\gamma^*_l,{\tilde{I}}_l^*]=\displaystyle\arg\max\limits_{\gamma_l,\tilde{I}_l}\; \frac{1}{{d_{ll}}^2}\times \frac{\log_2\big(1+\tfrac{c \gamma_l {d_{ll}}^{-\alpha}}{\tilde{I}_l}\big)}{\frac{3\sqrt{3}}{2}\times\Big[F^{-1}\big(\tfrac{c \gamma_l {d}_{ll}^{-\alpha}}{\tilde{I}_l}\big)\Big]^2}}\\[3pt]
                  \qquad\; \text{s. t.       }\; \displaystyle\frac{2\Gamma_c+g_a \gamma_l}{\log_2\big(1+\tfrac{c \gamma_l {d}_{ll}^{-\alpha}}{\tilde{I}_l}\big)}\leq \hat{E}_l,\, \gamma_l\times \tilde{I}_l=\lambda, \; \\ F(\frac{r_g}{d}) \geq \eta_{\min} \qquad \qquad \qquad \qquad \qquad \\
                    %\qquad\gamma_l\times \tilde{I}_l=\lambda \qquad \qquad \quad  \qquad\quad \;
  \end{aligned}
\end{equation}
%,\gamma^*_l\in[\gamma^{\text{min}}_l, \gamma^{\text{max}}_l], I^*_l\in[I^{\text{min}}_l, I^{\text{max}}_l]
where $\hat{E}_l$ is the maximum energy consumption per bit threshold at link $l$, and constant $\lambda$ should be chosen based on the feasible range of transmission power and interference bound of the links. We numerically solve (\ref{C4-18b}) using a brute-force search over discrete values of $\gamma_l$ and $\tilde{I}_l$.

\begin{figure}
  \centering
  \includegraphics[width=.5\textwidth, trim=2.75cm 10.7cm 6.5cm .5cm, clip=true]{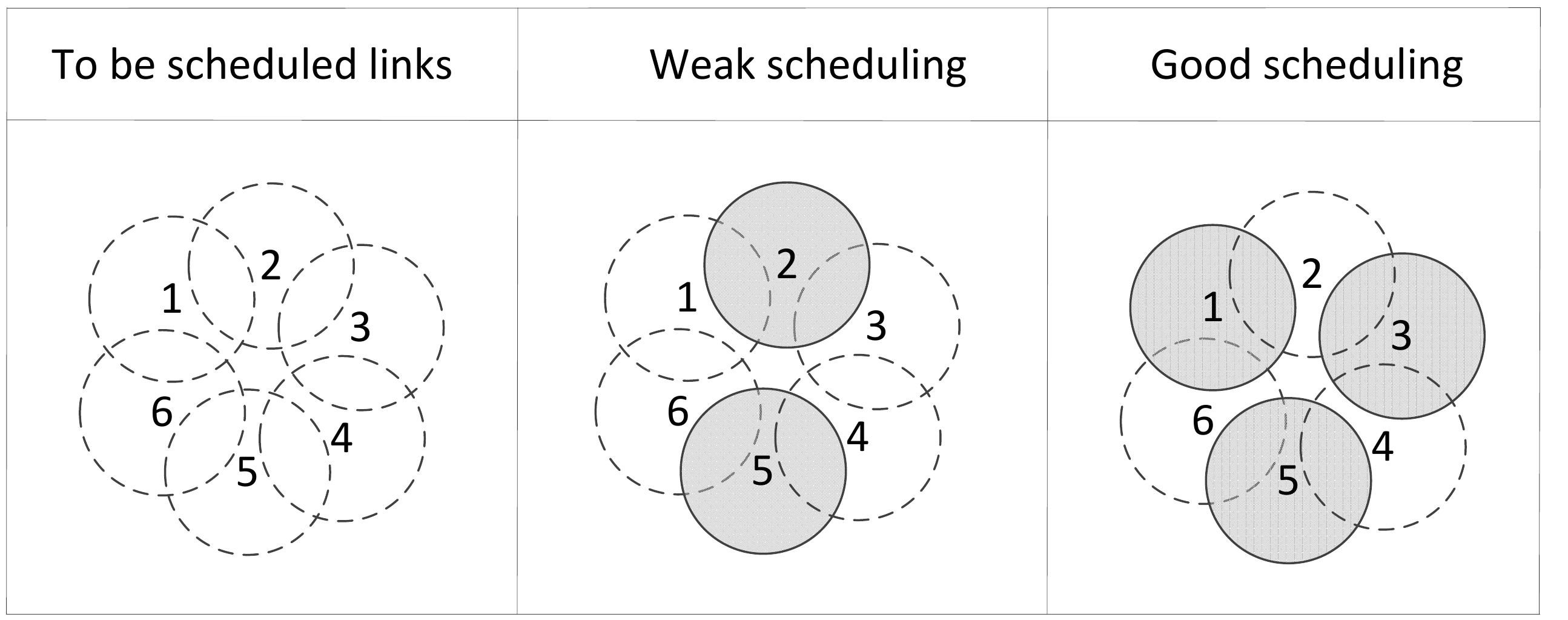}
  \caption{Weak link scheduling plan versus good link scheduling plan}\label{LinkSchFig}
\end{figure}

\subsection{Link scheduling}\label{LinkSch}
Given the transmission power and target interference of different links, concurrent links for transmissions should be properly determined such that the actual power of interference at the destination of scheduled links are close to their target interference power levels. For instance, consider the scheduling scenario illustrated in Figure \ref{LinkSchFig}. The first column shows six links to be scheduled. For simplicity of illustration, we use the circular areas to show link areas based on their transmission power and target interference power levels. Any two links can be scheduled simultaneously only if their circular areas do not overlap. The scheduled links are indicated by shaded circular areas in the second and third columns. The second column shows a weak scheduling plan in which only two links can be scheduled. A better scheduling plan is represented in the third column in which three links are scheduled by properly selecting the set of concurrent scheduled links. The better scheduling plan that schedules more concurrent links corresponds to the situation where the actual interference power levels are closer to the target interference power levels in the scheduled links, in comparison to the weak scheduling plan.

\begin{figure}[t]
  \centering
  \includegraphics[width=3.63in, trim=5cm .5cm 3.cm .5cm, clip=true]{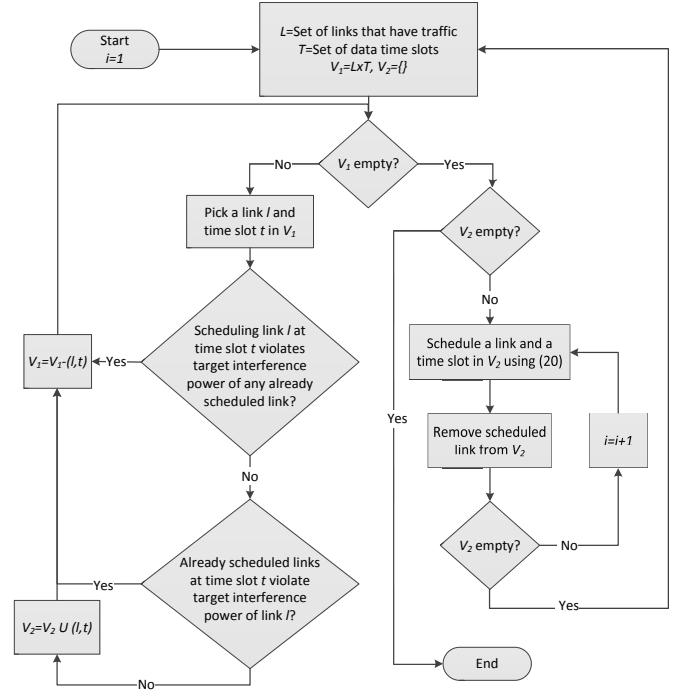}\\
  \caption{Flowchart operations of proposed link scheduling scheme.}
  \label{FlowChart}
\end{figure}

We consider a sequential link scheduling scheme to avoid high complexity. At each step, one link is scheduled for transmission at a time slot, which are opportunistically determined to have the interference power as close as possible to the target interference level at the scheduled destinations. Let $\bar{\boldsymbol{u}}^i={[u_{lt}^i]}_{L\times T}$ denote the scheduling matrix after step $i$, with $\bar{\boldsymbol{u}}^0={[0]}_{L\times T}$. The data rate of link $l$ up to sequential scheduling step $i$ is $R_{l}^i=\frac{1}{T}\sum_{t=1}^{T}\log_2\big(1+u_{lt}^i\gamma^*_{l}h_{ll}/\tilde{I}^*_l\big)$. Let $\hat{\gamma}_{lt}^i$ denote the maximum transmission power at the source node of link $l$ at slot $t$, which does not increase the interference power at any already scheduled link before step $i$ to more than its target interference power level. We have
\begin{equation}\label{C4-20}
 \hat{\gamma}_{lt}^i=\min_{k}\Big(\frac{\tilde{I}^*_{k}-\sum_{j\neq k}u_{jt}^{i-1}\gamma^*_j h_{jk}}{h_{lk}}\Big),\, \scalebox{.8}{$k\neq l, u_{kt}^{i-1}=1$}.
\end{equation}
Similarly, let $\hat{I}_{lt}^i$ denote the minimum possible target interference power for link $l$ at slot $t$ in the presence of already scheduled links before step $i$. We have
\begin{equation}\label{C4-21}
 \hat{I}_{lt}^i=\sum_{k}u_{kt}^{i-1}\gamma^*_k h_{kl}, \, \scalebox{.8}{$k\neq l$}.
\end{equation}
Thus, at step $i$, link $l$ can be scheduled at time slot $t$ if $\hat{\gamma}_{lt}^i \geq \gamma^*_l$ and $\hat{I}_{lt}^i \leq \tilde{I}^*_l$. The ratio $\hat{I}_{lt}^i / {\tilde{I}^*_l}$ indicates how close the target interference power and the actual interference power are at link $l$ at slot $t$ in $i^{\text{th}}$ step, while $\gamma^*_l / \hat{\gamma}_{lt}^i$ is the indication for the link closet to link $l$, after scheduling link $l$ at slot $t$ at $i^{\text{th}}$ step. Thus, at step $i$, we schedule link $l^i$ for time slot $t^i$ for highest product $\hat{I}_{lt}^i / {\tilde{I}^*_l} \times \gamma^*_l / \hat{\gamma}_{lt}^i$, given by
\begin{equation}\label{C4-23}
  \begin{array}{c}
  [l^i,t^i]=\arg\max\limits_{l,t}{\Big(\dfrac{\gamma^*_l}{\hat{\gamma}_{lt}^i}\times \dfrac{\hat{I}_{lt}^i}{\tilde{I}^*_l}\Big)}, \; \scalebox{.8}{$l\in\big\{1,2,...,L\big\},\, t\in\big\{1,2,...,T\big\}$}\\

           \text{   s. t.     }\,    \hat{\gamma}_{lt}^i \geq \gamma^*_l \text{,  } \hat{I}_{lt}^i \leq \tilde{I}^*_l, \, R_{l}^i\leq \hat{R}_l.
  \end{array}
\end{equation}
In each step, the solution of (\ref{C4-23}) can be calculated using a brute-force search over all links and time slots. For fairness, scheduling is performed in several rounds and in each round a link is scheduled at most once. The sequential scheduling steps in each round continue until every link is either scheduled once or cannot be scheduled. The scheduling rounds continue until no new link can be scheduled. Figure \ref{FlowChart} illustrates operations of the proposed link scheduling scheme.

\begin{figure}[t]
  \centering
  \includegraphics[width=2.40in, trim=8.5cm 4.8cm 7cm 4.8cm, clip=true]{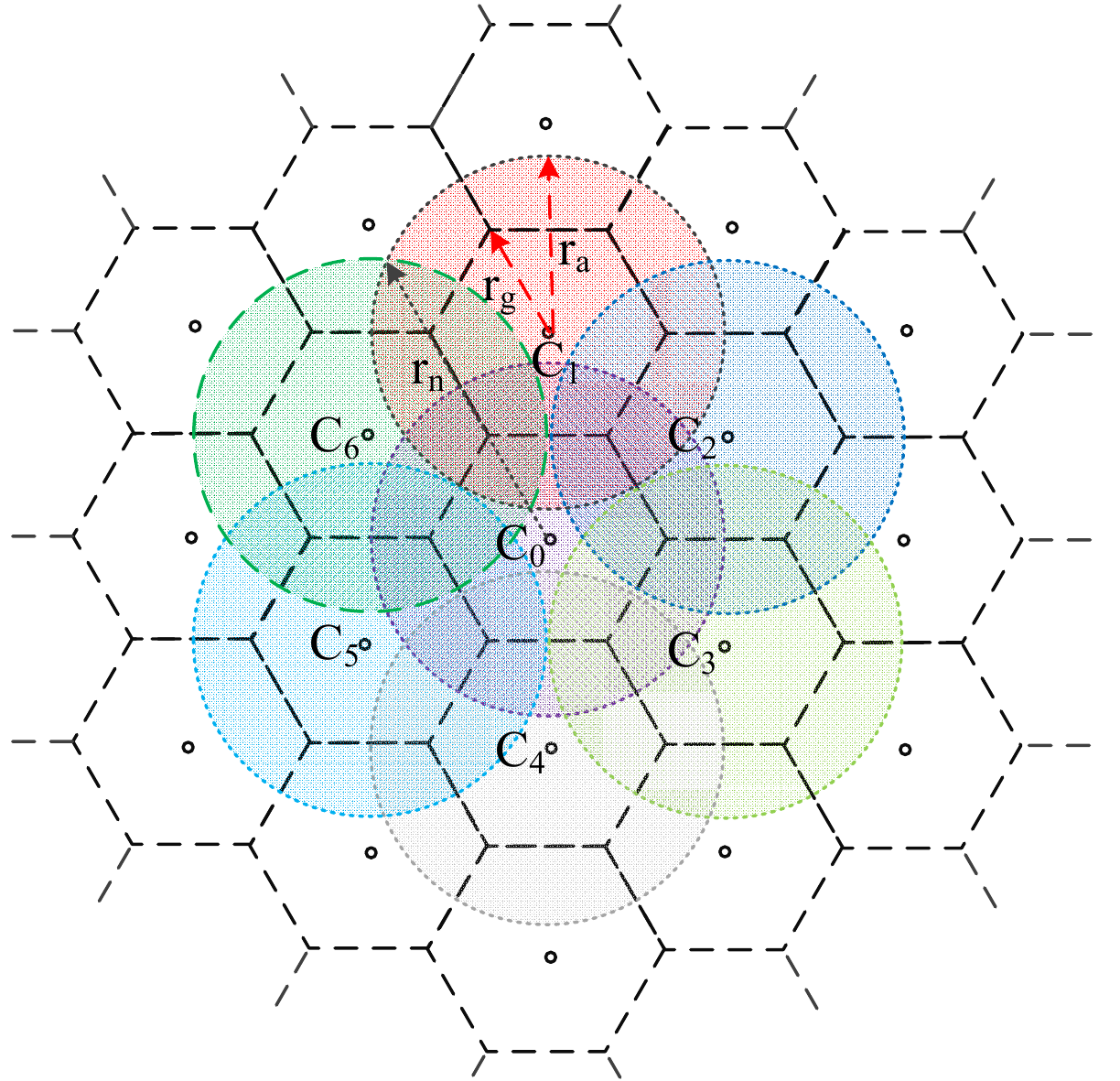}\\
  \caption{Partitioning the network area into hexagonal cells, where $C_i$, $i\in\{0,1,2,...\}$, denotes the coordinator of cell $i$. A circular area centred at each coordinator denotes the location area of the nodes that their scheduling information is broadcasted by the coordinator ($r_a=1.5r_g$).}
  \label{InfR-n1}
  \vspace{.5cm}
  \centering
  \includegraphics[width=.46\textwidth, trim=.6cm 13.2cm .9cm 5.6cm, clip=true]{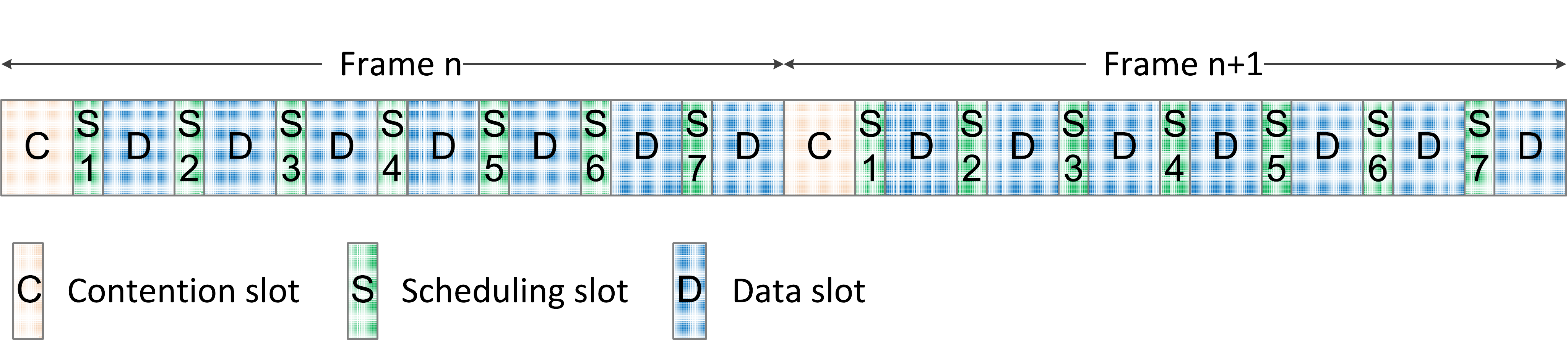}\\
  \caption{Structure of one frame of the proposed MAC framework.}
  \label{FrameStructure-n1}
\end{figure}

\subsection{Distributed scheduling}
It is desired to have a distributed implementation of the proposed scheduling and transmission power control scheme, based on only local network information. According to (\ref{C4-18b}), the transmission power and target interference power can be determined independently at each link. In Subsection \ref{LinkSch}, links are scheduled sequentially based on the information of already scheduled links using (\ref{C4-23}). However, the information of local scheduled links is the most relevant information to schedule links for transmission, because the power of interference decreases exponentially with distance. The power of interference at the destination node of link $l$ at time slot $t$ caused by source nodes of the scheduled links at distance farther than $d_0$ ($>0$) is
\begin{equation}\label{C4-25}
\sum_{\scalebox{.8}{$k\neq l, d_{kl} > d_0$}}u_{kt}\gamma_{kt}h_{kl}\leq c_0 \gamma_{\max}{d_0}^{-\alpha}\triangleq I_0
\end{equation}
where $c_0$ is a constant and $\gamma_{\max}$ denotes the maximum transmission power level. Thus, using only the information of scheduled local links within distance $d_0$ and $I_0$, we can estimate the power of interference at a link to calculate (\ref{C4-20}) and (\ref{C4-21}) that are required for the link scheduling scheme in (\ref{C4-23}). As an example, consider scheduling of link $l$ at time slot $t$ when two other links are already scheduled at time slot $t$ within distance $d_0$ with transmission power and target interference levels $\gamma^*_1$, $\tilde{I}^*_1$ and $\gamma^*_2$, $\tilde{I}^*_2$, respectively. We have $\hat{\gamma}_{lt}^i=\min\Big(\frac{\tilde{I}^*_{1}-\gamma^*_2 h_{21}-I_0}{h_{l1}}, \frac{\tilde{I}^*_{2}-\gamma^*_1 h_{12}-I_0}{h_{l2}}\Big)$ and $\hat{I}_{lt}^i=\gamma^*_1 h_{1l}+\gamma^*_1 h_{1l}+I_0$.

To coordinate distributed link scheduling, we employ a set of coordinator nodes distributed over the network area to collect and exchange local network information and to periodically schedule links in a distributed manner. In the following, we describe the proposed MAC framework (which is based on our scheme proposed in \cite{Kamal3}) to coordinate link scheduling based on source node transmission requests. The network coverage area is partitioned into hexagonal cells as shown in Figure \ref{InfR-n1}. The distance $r_g$ between the center and a vertex of a cell is chosen such that $r_g\geq d_{\max}$. Therefore, the destination node of each source node is either in the same cell or an adjacent cell. A coordinator node is placed at the center of each cell to coordinate all transmissions for nodes inside the cell. Figure \ref{FrameStructure-n1} shows the frame structure. Each frame consists of three types of time slots:
\begin{enumerate}
  \item \begin{it}Contention slots:\end{it} During contention slots, the source nodes that want to initiate a transmission contend with each other using a truncated CSMA MAC scheme to send a request packet to the cell coordinators. If the number of contention slots is too small, nodes may not have enough time to transmit requests to initiate transmissions. On the other hand, assigning a large number of slots as contention slots decreases the number of data transmission slots which reduces network throughput. We have presented a mathematical model in \cite{Kamal3} to determine the number contention slots in coordinators based on traffic load condition;
  \item \begin{it}Scheduling slots:\end{it} Each coordinator node has a scheduling time slot in every frame, in which it broadcasts a scheduling packet to coordinate all transmissions in its vicinity;
  \item \begin{it}Data slots:\end{it} Data packet transmissions are performed during contention-free data slots as scheduled by the coordinators. A link transmits one data packet during a data slot that is scheduled for transmission.
\end{enumerate}
A coordinator node maintains the following information about each link in its vicinity:
\begin{enumerate}
  \item The source and destination locations;
  \item The transmission power and target interference level;
  \item The set of future data slots that it is scheduled;
  \item The amount of data that it has for transmission.
\end{enumerate}
A coordinator receives transmission requests from source nodes during contention slots. Also, a coordinator receives the information of scheduled links for the future data slots by overhearing scheduling packets of adjacent coordinators during scheduling slots. The scheduling packets of a coordinator contains the information of all future scheduled data transmissions for every node within distance $r_a$ ($\geq r_g$) from the coordinator\footnote{An interesting future work is to consider increasing $r_a$ to enlarge the area that each coordinator acquires the information of scheduled transmissions, and thus to enable scheduling single-hop transmission between two nodes with distance larger than $r_g$ (i.e., when  single-hop source and destination nodes are not in a cell or adjacent cells).}. Figure \ref{InfR-n1} shows the area centred at a coordinator where the coordinator obtains the information of scheduled transmissions by overhearing scheduling packets of adjacent coordinators. According to Figure \ref{InfR-n1}, a coordinator node acquires the information of scheduled transmissions within distance $r_n=1.5r_g+\sqrt{{r_a}^2-0.75{r_g}^2}$ and for each link, depending on the destination node's location in the cell, we have $d_0\in[r_n-r_g,r_n]$. Based on the source node requests for transmission and the information of already scheduled links, each coordinator periodically schedules data transmissions for every link with the destination inside its cell in the future data slots before its own subsequent scheduling slot. A coordinator node schedules links for transmission according to the proposed link scheduling scheme in Subsection \ref{LinkSch} (with the consideration of already scheduled links by adjacent coordinators) and broadcasts a scheduling packet in its scheduling slot to announce the scheduling information to nodes inside its cell and to its adjacent coordinators. The scheduled links perform data transmissions during data time slots as scheduled by cell coordinators and announced during scheduling slots. Every node puts its radio interface in the sleep mode when it is not transmiting/receiving a scheduling, data or request packet to save energy.

\begin{figure*}
\centering
\subfigure[]{\includegraphics[width=2.31in, trim=.05cm 1.5cm 1cm 1.4cm, clip=true]{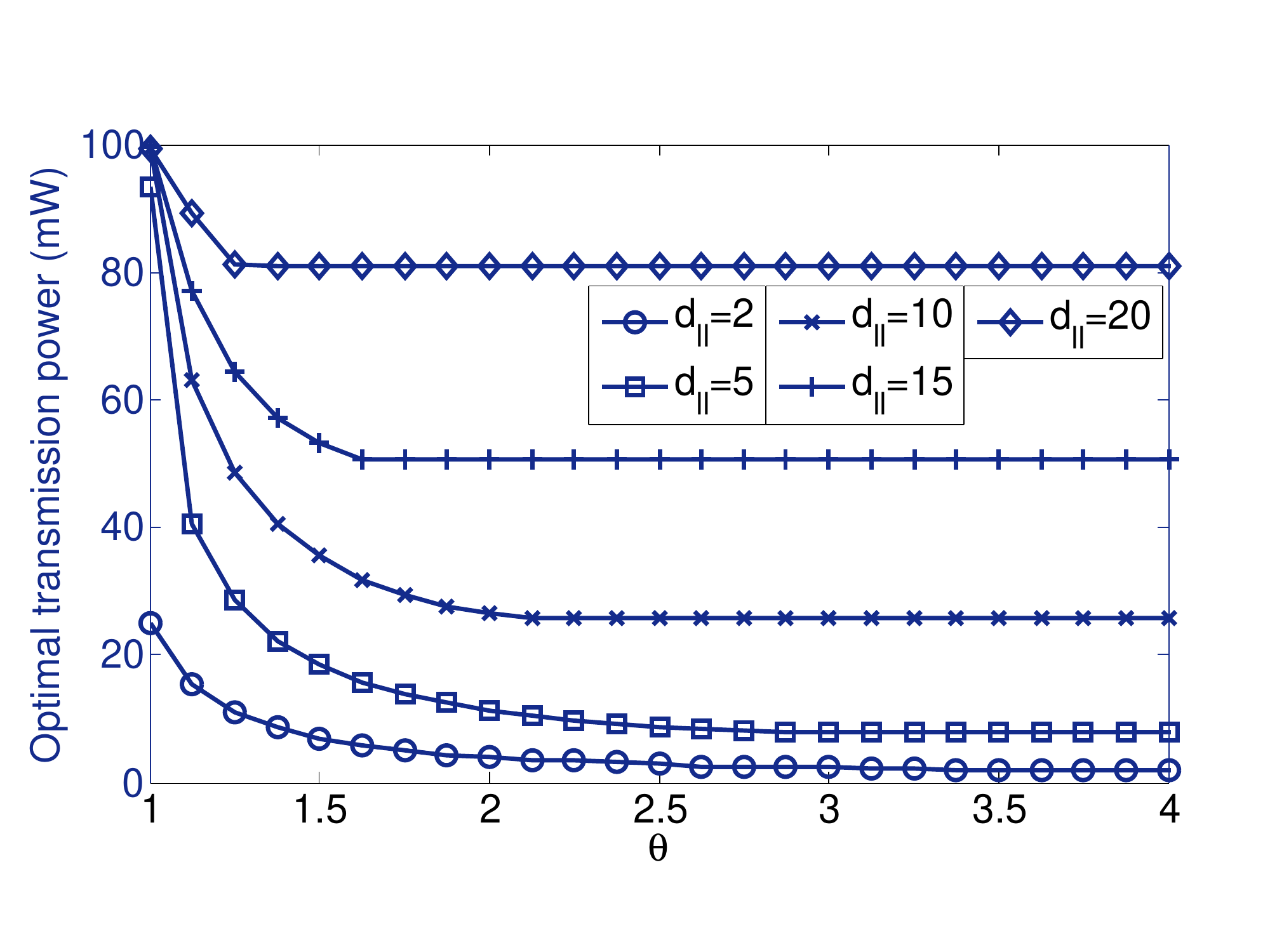}\label{Num_OPTTrP}}
\subfigure[]{\includegraphics[width=2.31in, trim=.05cm 1.5cm 1cm 1.4cm, clip=true]{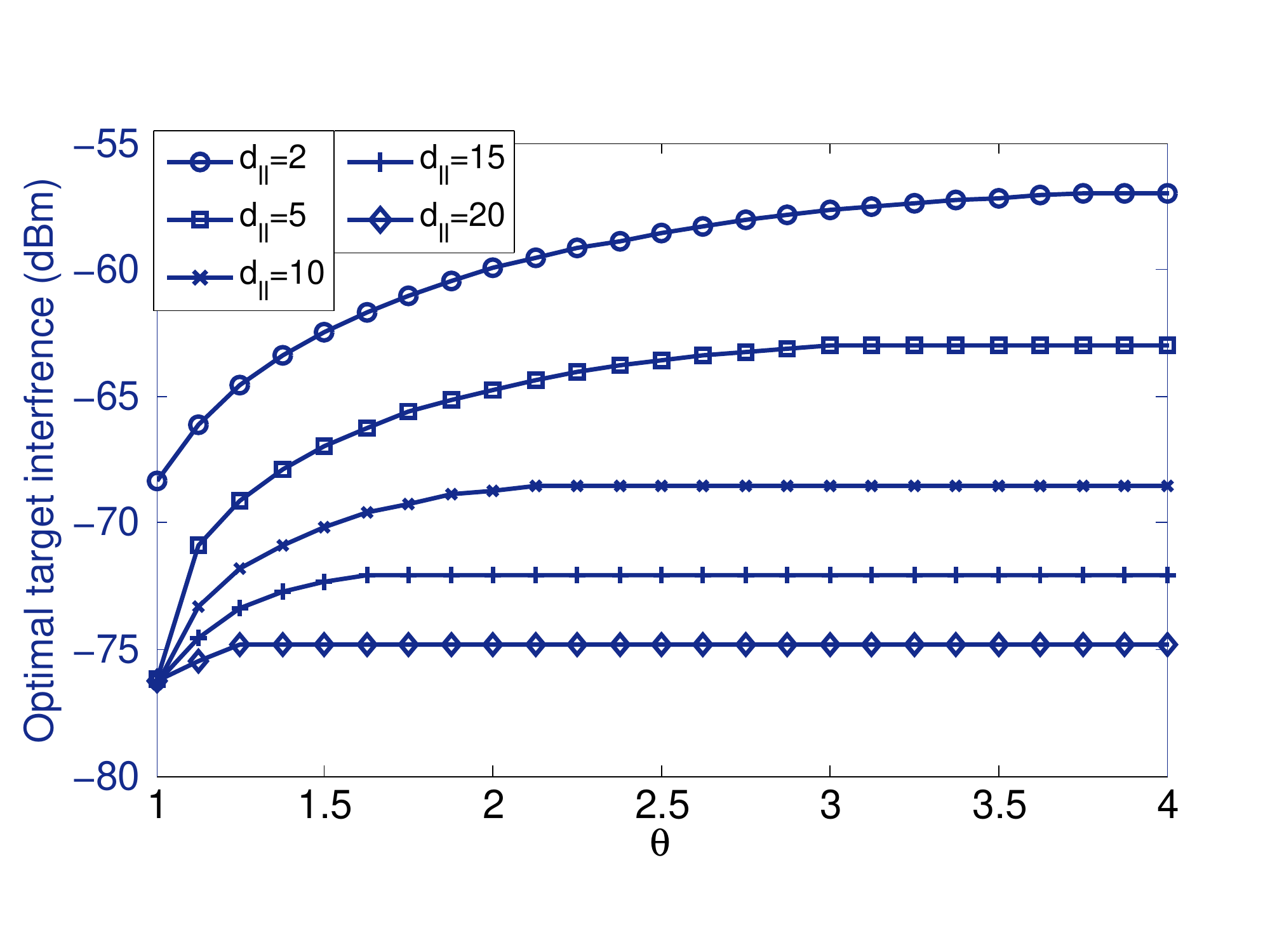}\label{Num_OPTInB}}
\subfigure[]{\includegraphics[width=2.31in, trim=.05cm 1.35cm 1cm 1.4cm, clip=true]{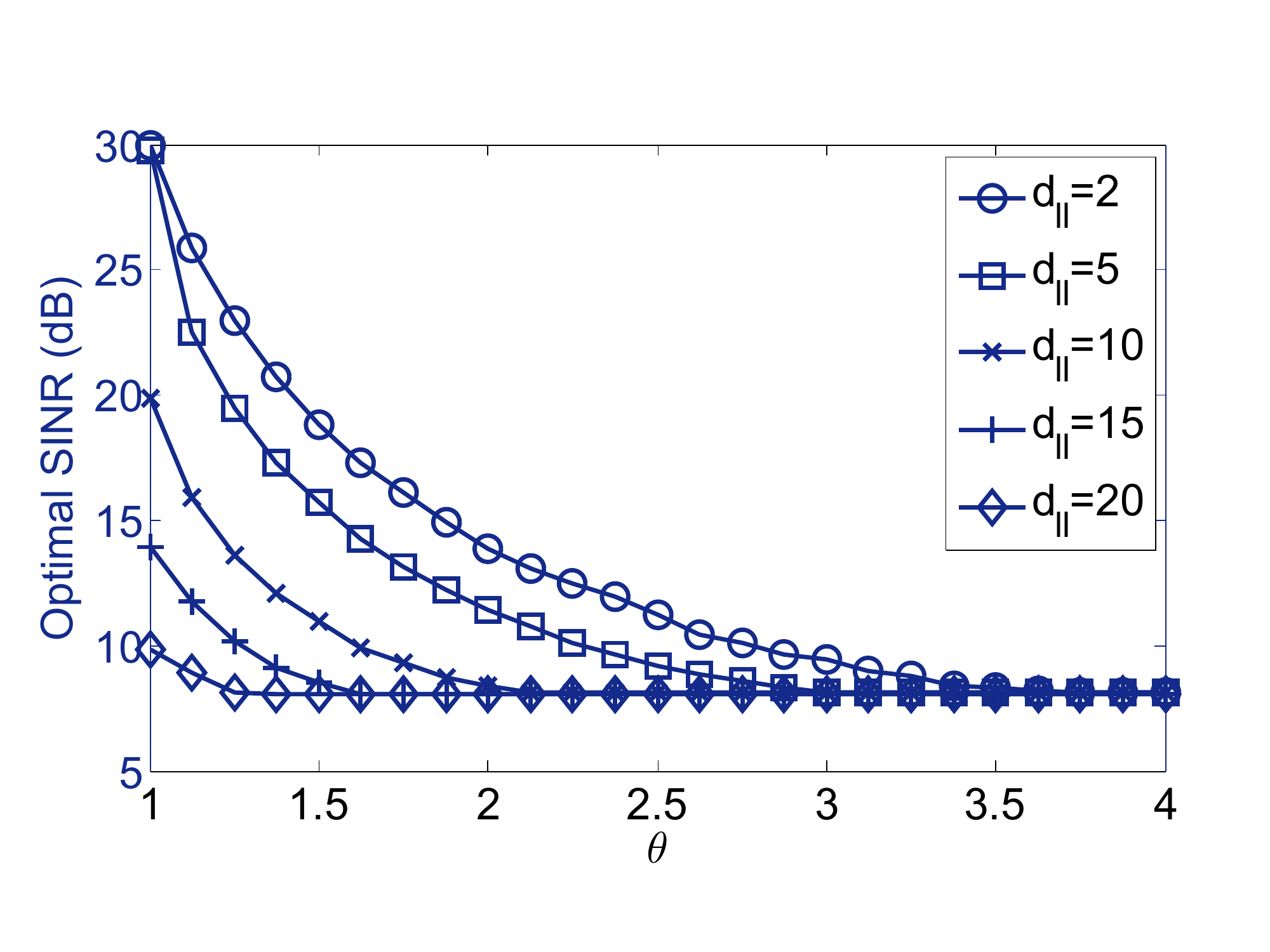}
\label{Num_OPTSINR}}
\caption{Optimal transmission power, target interference level and SINR versus $\theta$ as link distance $d_{ll}$ (m) varies.}\label{Num_OPT}
\end{figure*}

\begin{figure*}
\centering
\begin{minipage}{.64\textwidth}
\subfigure[]{\includegraphics[width=2.31in, trim=.05cm 1.5cm 1cm 1.4cm, clip=true]{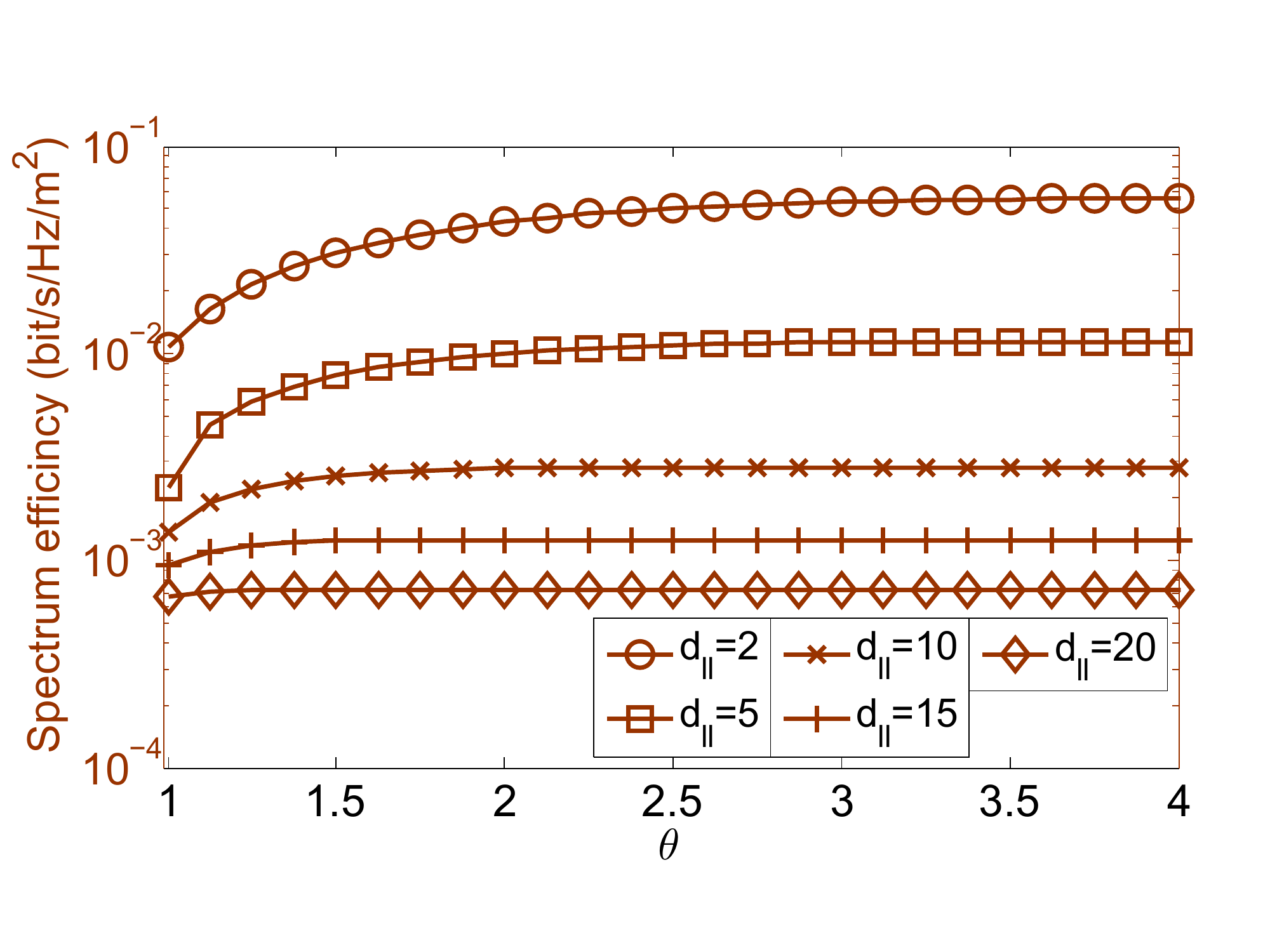}\label{Num_OPTSEd2}}
\subfigure[]{\includegraphics[width=2.31in, trim=.05cm 1.5cm 1cm 1.4cm, clip=true]{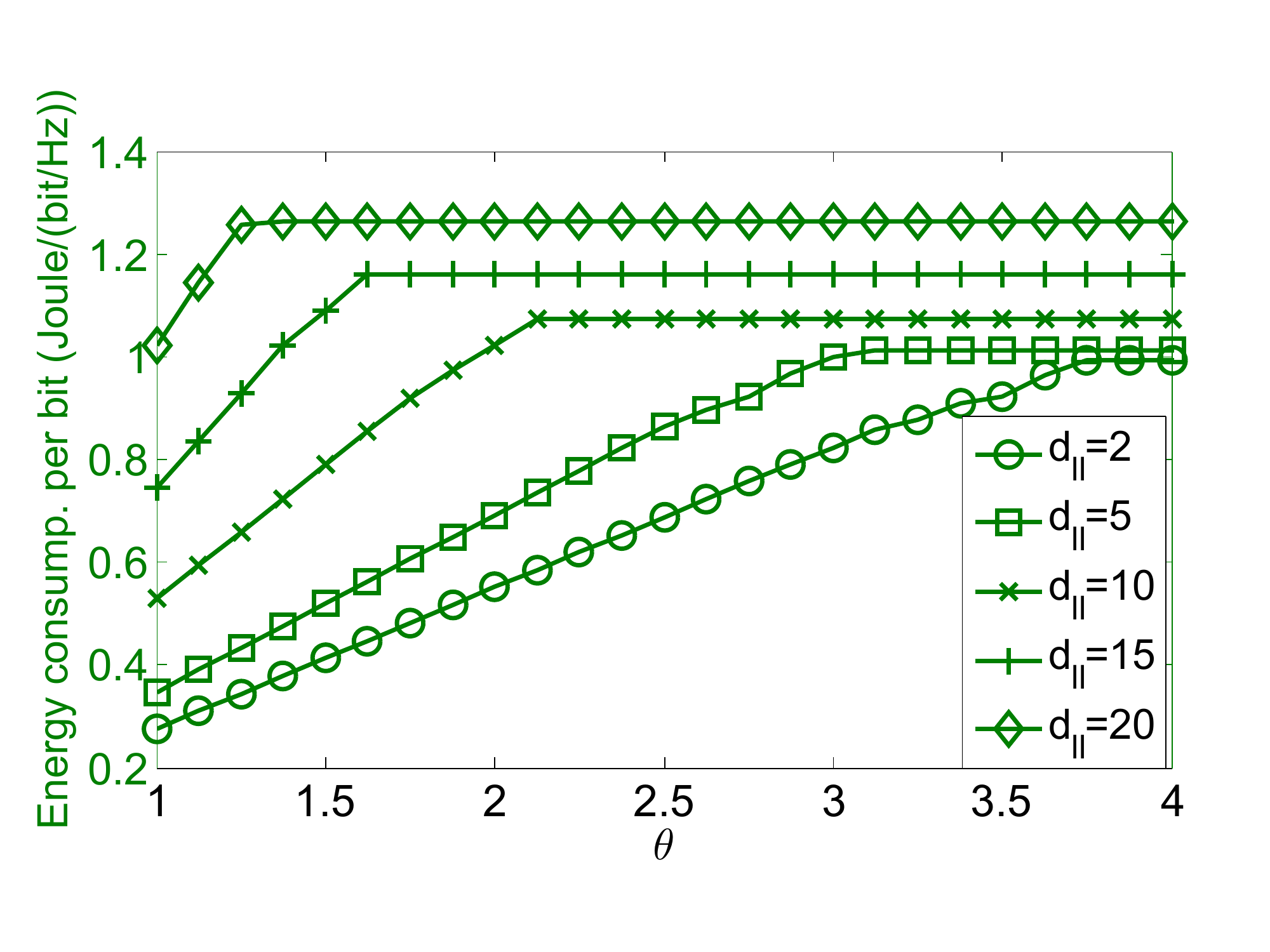}\label{Num_OPTEC}}
\caption{Asymptotic spectrum efficiency and energy consumption per bit versus $\theta$ as link distance $d_{ll}$ (m) varies. } \label{Num_OPTTSE}
\end{minipage}\hspace{.2cm}
\begin{minipage}{.32\textwidth}
\includegraphics[width=2.35in, trim=.13cm 2.7cm .25cm 2.7cm, clip=true]{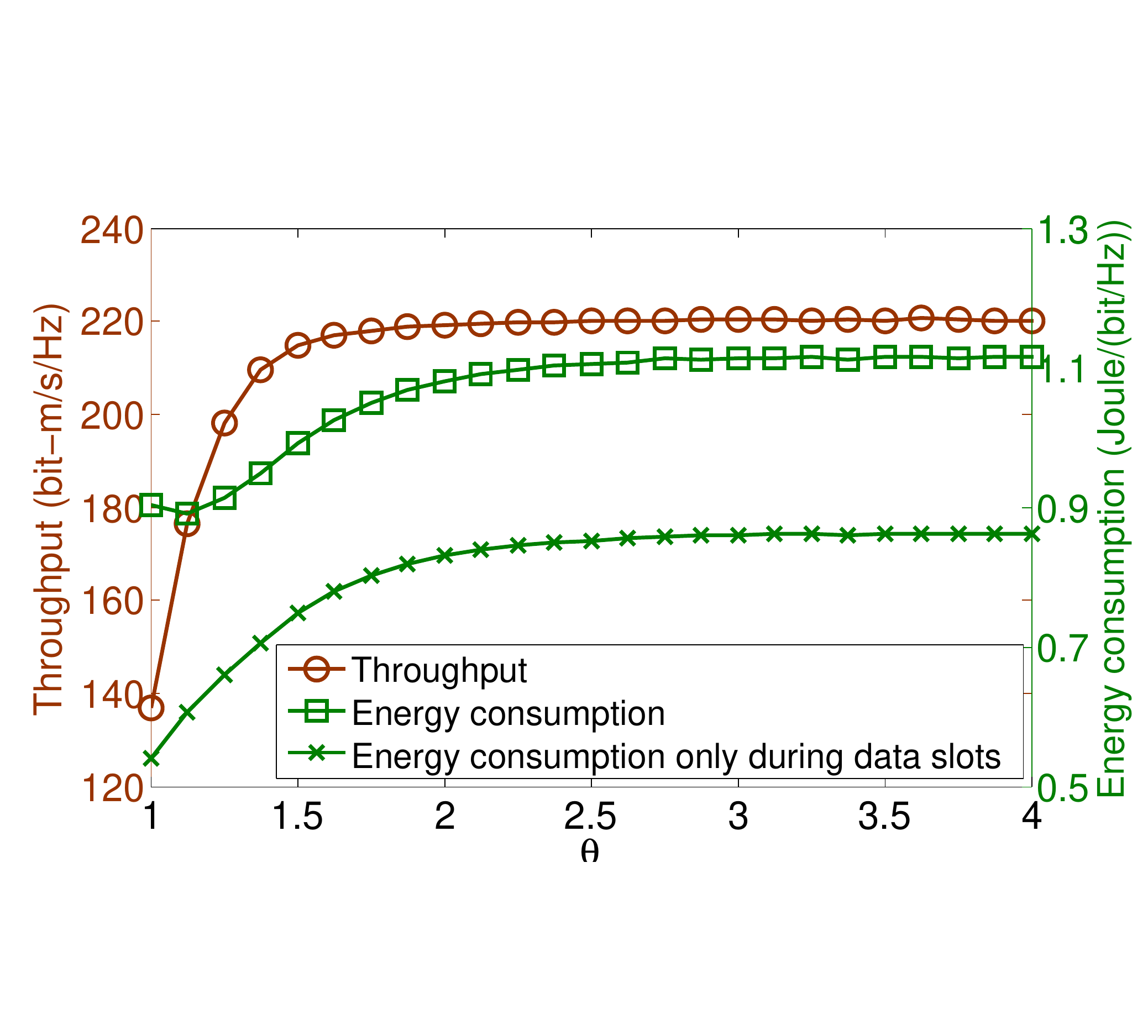}
\caption{Throughput and energy consumption of proposed scheme as $\theta$ varies with saturated data traffic at all nodes ($N=400$).} \label{TBIT1C4vsTEC4-TBIT0}
\end{minipage}
\end{figure*}

\renewcommand{\arraystretch}{1}
\begin{table}[tp]
    \begin{center}
    \caption{Simulation Parameters}
     \label{C4-Parameters}
        \begin{tabular}{l|l||l|l}
          \hline
          \hline
          Parameter & Value & Parameter & Value \\
           \hline
          Frame length &  100 ms & $c$ & 0.0001\\
          Data time slot &  1 ms & $\alpha$ & 3.4\\
          Scheduling time slot &  1 ms & $d_{\max}$ & 20 m\\
          Contention time slot &  1 ms & Bandwidth & 2 MHz\\
          Number of data slots & 90 &  $\gamma_{\max}$ & 100 mW\\
          Number of scheduling slots & 7 & $\gamma_{\min}$ & 1 mW\\
          Number of contention slots & 3 & $I_{\max}$ & -45 dB\\
          Data packet length &  1 ms & $I_{\min}$ & -80 dB\\
          Scheduling packet length &  1 ms & $\eta_{\max}$  & 30 dB\\
          Beacon interval &  100 ms & $\eta_{\min}$ & 6 dB\\
          ATIM size & 224 bits &  $R_s$ & 6 Mbps\\
          ATIM-ACK size &  112 bits & $\Gamma_c$ & 1.25 W\\
          Mini-slot &      20 $\mu$s & $g_a$ & 10\\
          SIFS      &      10 $\mu$s & $\Gamma_0$ & 0 W  \\
          PHY preamble & 72 $\mu$s  & $CW_{\min}$ &     15  \\
          Scheduling size for a transmission &   200  bits  & $CW_{\max}$ &    1023   \\
          Request packet size & 160 bits  & &\\
          \hline
          \hline
        \end{tabular}
    \end{center}
\end{table}
\renewcommand{\arraystretch}{1}

\section{Numerical Results} \label{SimRes}
Consider area of 19 hexagonal cells as illustrated in Figure \ref{InfR-n1}. There are $N$ nodes randomly distributed over the area. The destination node of each link is randomly selected from the nodes within distance $d_{\max}$ from the source node. The ranges of feasible transmission power level, target interference power level and SINR value for a link are provided in Table \ref{C4-Parameters} based on IEEE 802.11 standard \cite{Standard}. We set the energy consumption per bit constraint, $\hat{E}_{l}=\theta \times \min E_{l}$ for every link $l$, where $\theta \geq 1$. Thus, $\theta=1$ corresponds to setting transmission power and target interference power for lowest energy consumption per bit in each link, while as $\theta$ increases, the energy consumption constraint is relaxed and the transmission power and target interference of a link are determined based on the values that provide highest asymptotic spectrum efficiency. Figure \ref{Num_OPT} shows the optimal transmission power, target interference level and SINR of a link versus $\theta$ as the link distance varies. The corresponding asymptotic spectrum efficiency and the energy consumption per bit are depicted in Figures \ref{Num_OPTSEd2} and \ref{Num_OPTEC} respectively. Figure \ref{Num_OPTInB} shows that the calculated optimal target inference level is always much larger than the thermal noise power level\footnote{Thermal noise power is about -101 dBm per 20 MHz.}, which conforms with the assumption of interference dominated network used in Section \ref{AsymOpt}. According to Figure \ref{Num_OPTSINR}, the SINR is set to the highest value for a link when the objective is to minimize energy consumption per bit (i.e., $\theta=1$). However, the optimal SINR value to maximize the asymptotic spectrum efficiency when the energy consumption constraint is weakened is always about 8 dB, independent of the link distance.

\begin{figure*}
\centering
\begin{minipage}{.64\textwidth}
\subfigure[$N=400$, $\theta=\infty$]{\includegraphics[width=2.3in, trim=.13cm 1.1cm .8cm 1.4cm, clip=true]{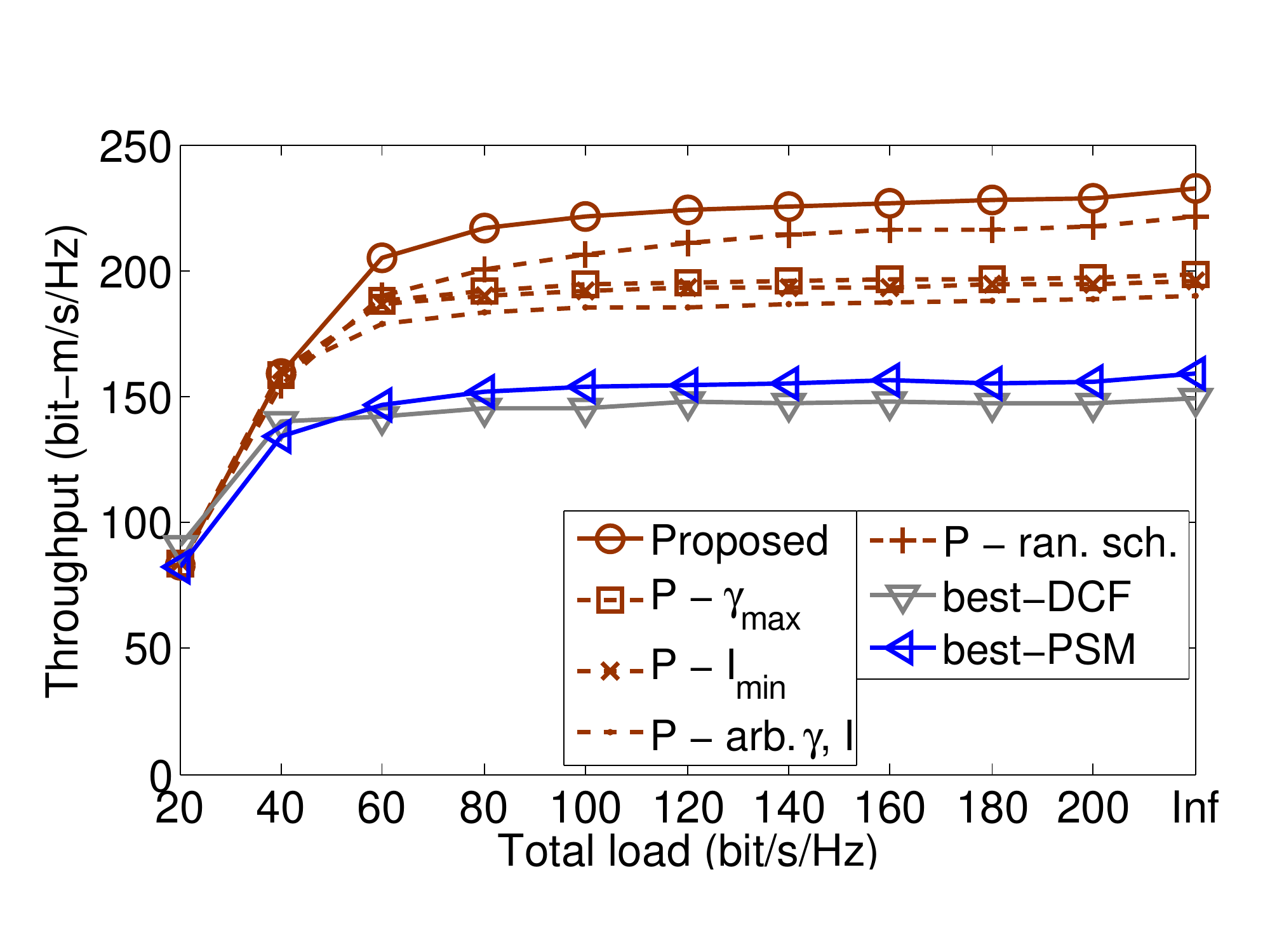}
\label{TBIT1}}
\subfigure[Saturated traffic load, $\theta=\infty$]{\includegraphics[width=2.32in, trim=.13cm 1.4cm .5cm 1.5cm, clip=true]{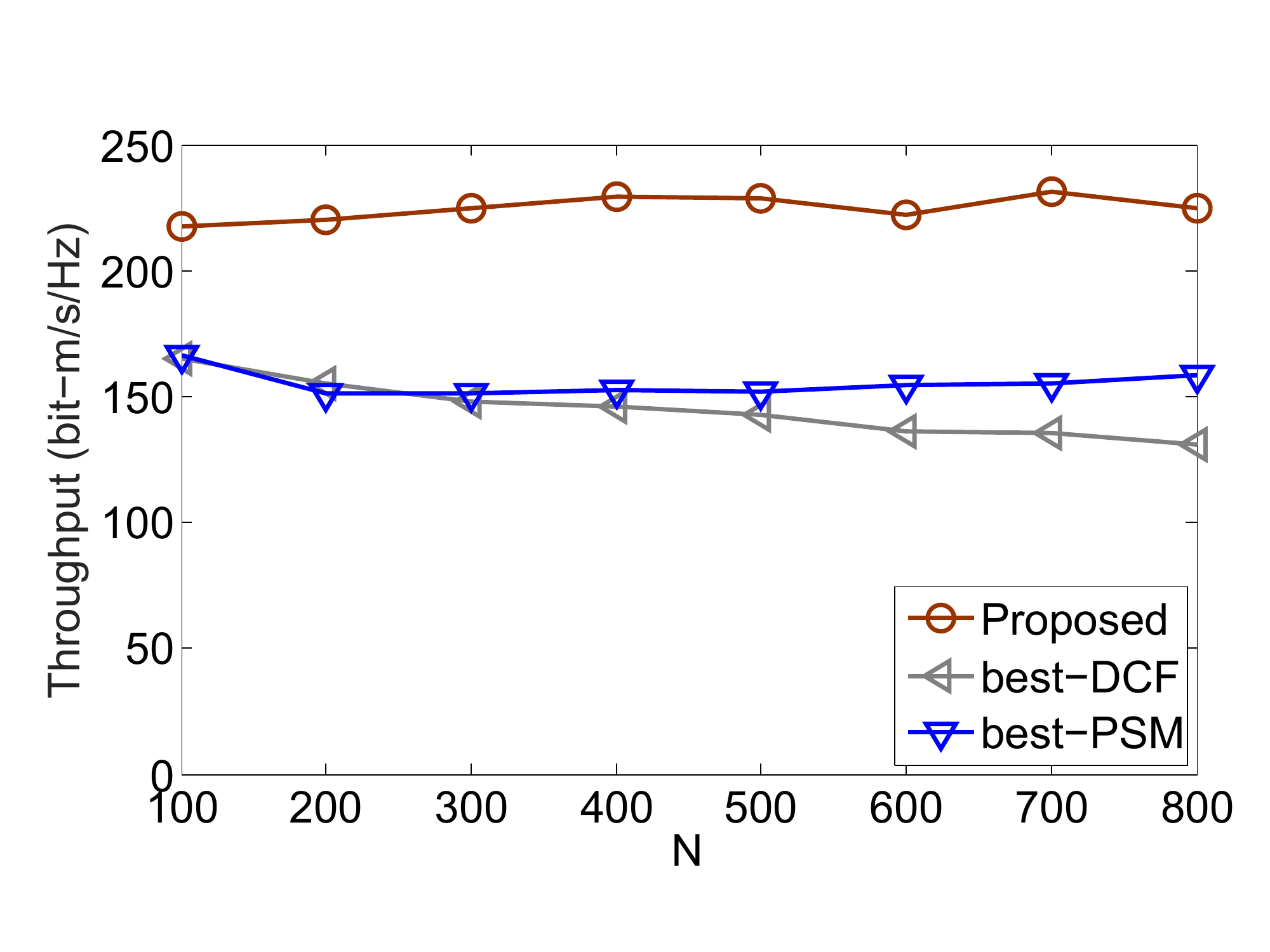}\label{TBIT1N}}
\caption{Throughput of different schemes as network traffic load and number of nodes vary.} \label{TBIT1-A}
\vspace{.15cm}
\end{minipage}\hspace{.3cm}
\begin{minipage}{.32\textwidth}
\includegraphics[width=2.31in, trim=.2cm 1.5cm 1cm 1.4cm, clip=true]{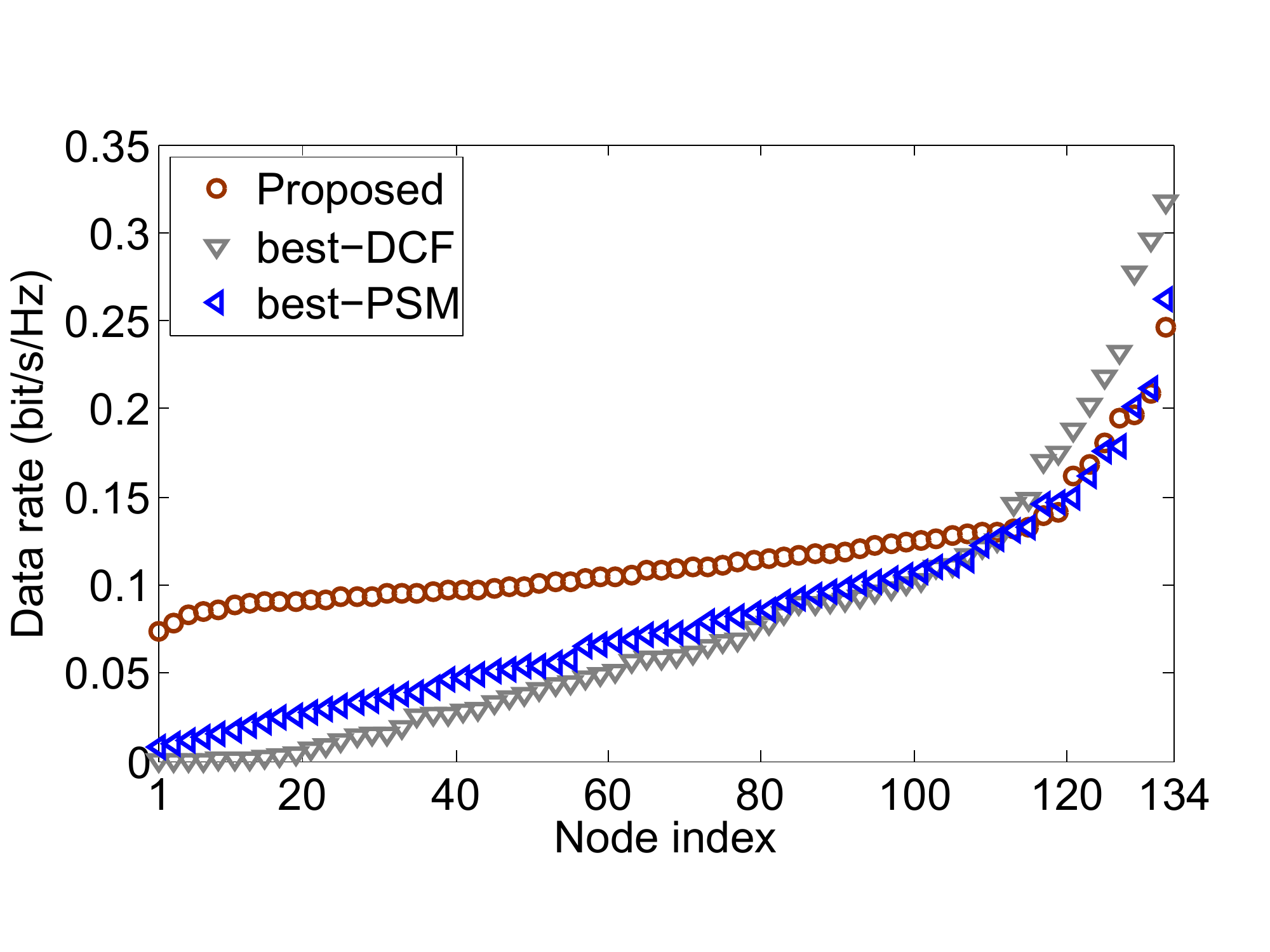}
\caption{Data transmission rate of the nodes using different schemes (Saturated data traffic, $N=400$, $\theta=\infty$).} \label{TBIT0peNode}
\vspace{.2cm}
\end{minipage}
\begin{minipage}{.64\textwidth}
\subfigure[$N=400$, $\theta=\infty$]{\includegraphics[width=2.3in, trim=.2cm 1.4cm 1cm 1.5cm, clip=true]{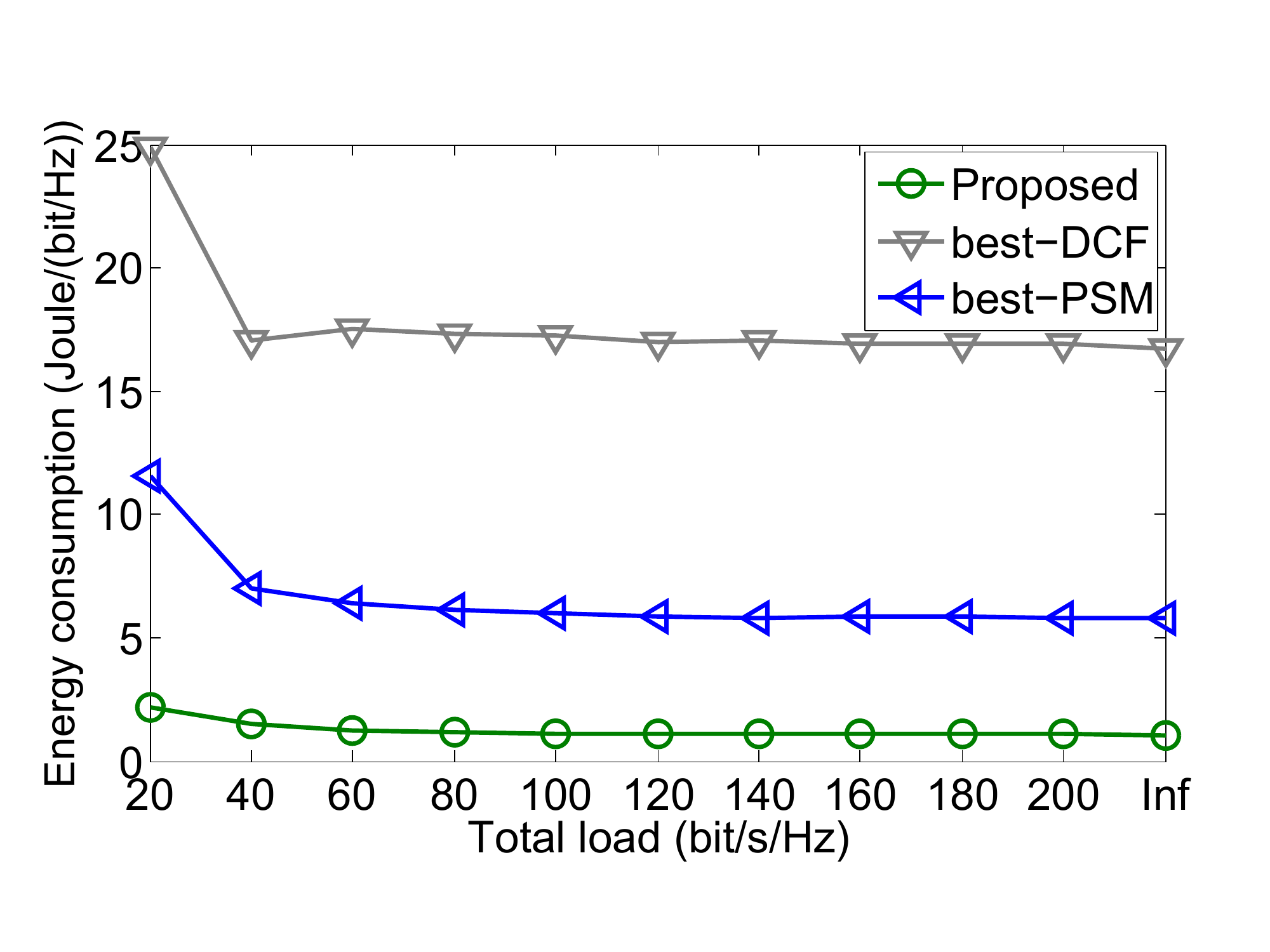}\label{TEC-TBIT0}}
\subfigure[Saturated traffic load, $\theta=\infty$]{\includegraphics[width=2.32in, trim=.2cm 1.5cm 1cm 1.4cm, clip=true]{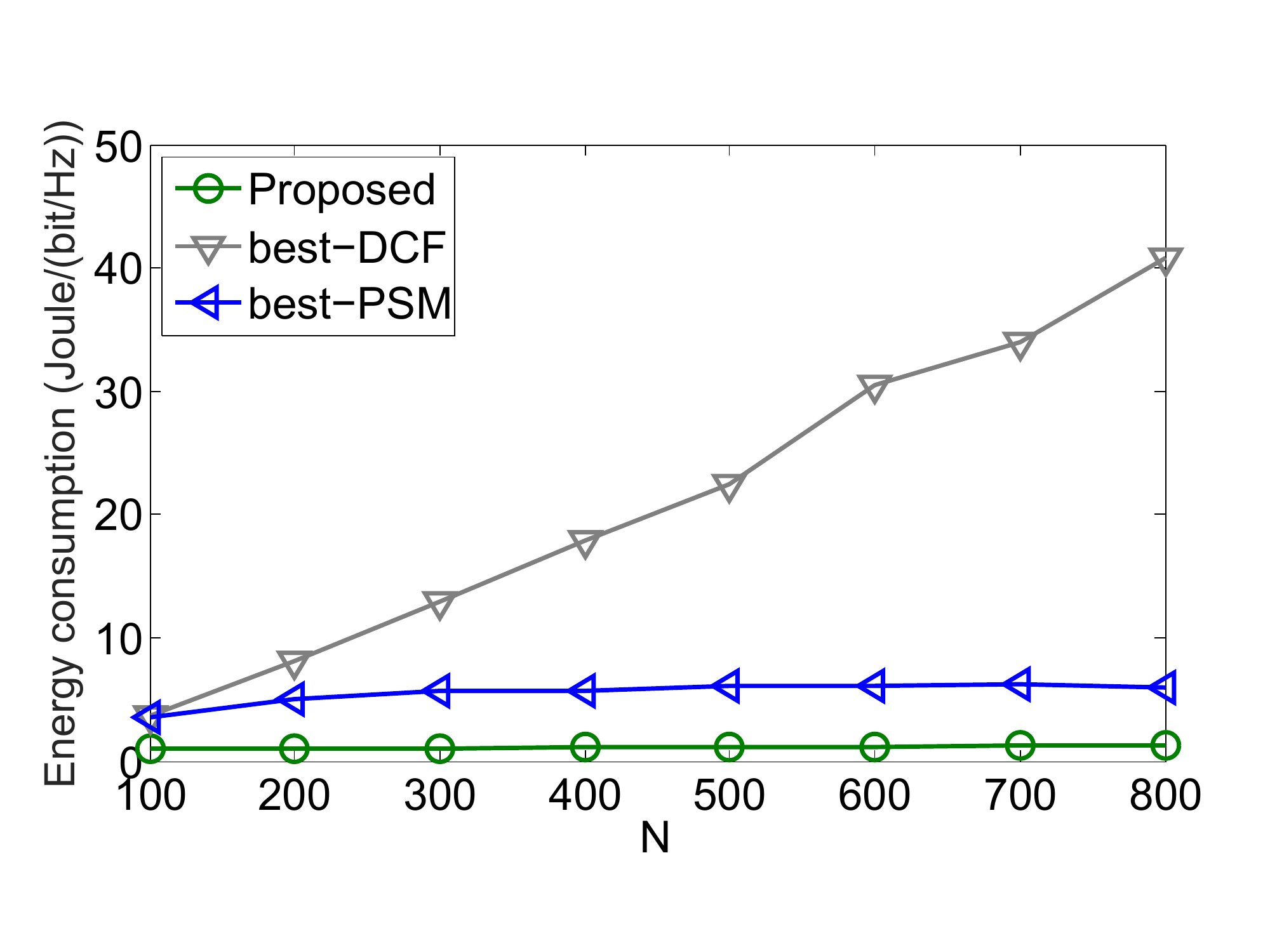}
\label{TEC-TBIT0N}}
\caption{Energy consumption of different schemes as network traffic load and number of nodes vary.}\label{TEC-TBIT0-A}
\end{minipage}\hspace{.2cm}
\begin{minipage}{.32\textwidth}
\includegraphics[width=2.31in, trim=.2cm 1.5cm .5cm 1.4cm, clip=true]{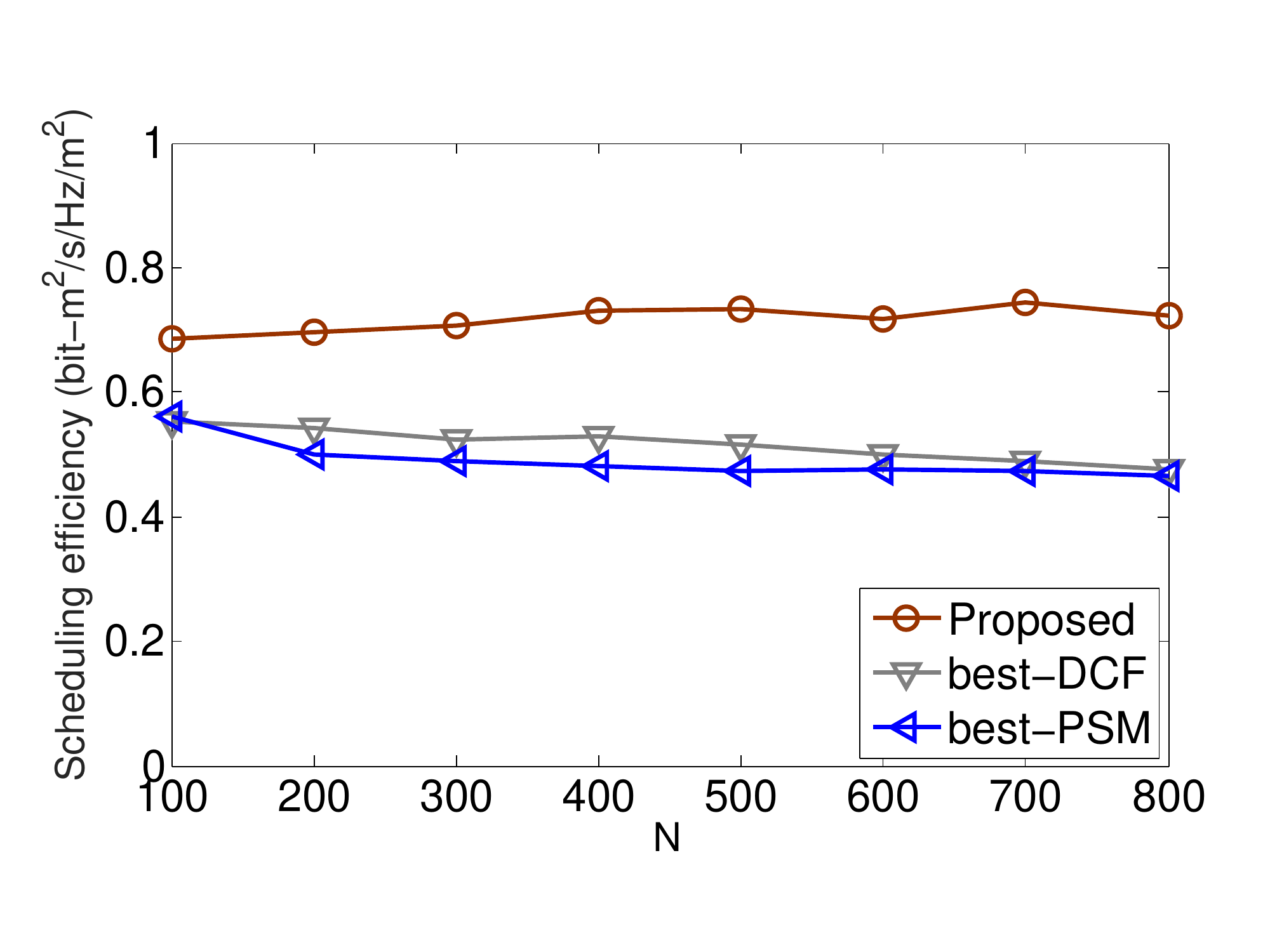}
  \caption{Scheduling efficiency of different schemes as number of nodes varies (Saturated data traffic, $\theta=\infty$).}\label{GSECHN}
\end{minipage}
\end{figure*}

We evaluate the performance of our proposed scheduling and transmission power control scheme via simulation. The following metrics are used as performance measure to compare different schemes:
\begin{enumerate}
  \item \begin{it}Throughput:\end{it} Throughput is defined as the summation of all transmitted data bits per second, weighted by the transmitted distance \cite{Gupta-Transpoty};
  \item \begin{it}Energy consumption:\end{it} Energy consumption is defined as the ratio of total energy consumed in the nodes to the total number of transmitted data bits. Similar metrics are also used in \cite{Kamal3, DPSM, Kamal2, TMMAC,Kamal};
  \item \begin{it}Scheduling efficiency:\end{it} According to (\ref{C4-5}), the spectrum efficiency for transmission distance $d$ is bounded by $\tilde{R}=1/d^2\times\max{G(\cdot)}$. Thus, the summation of all transmitted data bits per second, weighted by the second power of the transmitted distance, $\sum_{l} R_l d_{ll}^2\leq \max{G(\cdot)}\times A$, where $A$ denotes the area size and the equality holds under asymptotic optimal scheduling and transmission power control. Therefore, we define scheduling efficiency as the ratio $\sum_{l} R_l d_{ll}^2/(\max{G(\cdot)}\times A)$.
\end{enumerate}
The performance metrics are evaluated based on the transmitted data and energy consumption of the nodes in an inner region of the network area to eliminate edge effects. Links with source nodes located inside the 7 central hexagonal cells (of the 19 hexagonal cells) in Figure \ref{InfR-n1} and all coordinator nodes inside this area are considered in evaluating the performance metrics\footnote{This is because nodes at the edge of simulated network area experience less interference from their adjacent nodes, as the simulated area is bounded. Therefore, performance of nodes in the inner part of the simulated area should be considered to evaluate the actual network performance.}. We compare the performance of our proposed scheme with IEEE 802.11 DCF MAC with and without power saving and with optimized transmission power levels and carrier sensing threshold based on the analysis provided in \cite{Kim,Kim2}. Also, we examine the effectiveness of each strategy that we use for determining transmission power and target interference power levels and for link scheduling by evaluating the throughput without the strategy. The compared schemes are as follows:
\begin{enumerate}
  \item The proposed scheme, denoted by ``Proposed'';
  \item ``P - $\gamma_{\max}$'', ``P - $I_{\min}$'' and ``P - arb. $\gamma$, $I$'',  representing proposed scheme when the product of transmission power and target interference power is not maintained at a fixed value, but respectively the transmission power is set to the maximum value, the target interference level is set to the minimum value and the transmission power and target interference level are chosen arbitrary;
  \item ``P - ran. sch.'', representing the proposed scheme when the link scheduling by coordinators at each scheduling step is not according to the link scheduling algorithm described by (\ref{C4-23}), instead a link and a data slot are randomly selected from the set of links and slots that can be scheduled;
  \item ``best-DCF'' and ``best-PSM'', representing the DCF MAC of IEEE 802.11 in ad hoc mode without and with power saving mode respectively, with optimized transmission power levels and carrier sensing threshold based on the analysis provided in \cite{Kim} and with optimized ATIM window size\footnote{The ATIM window size in power saving mode is optimized for highest total network throughput using a brute force simulation of IEEE 802.11 DCF MAC with different ATIM window sizes.}.
\end{enumerate}
In each scheme, all control and signaling packets are transmitted using signaling rate $R_s$, which requires minimum SINR $\eta_{\min}$ during entire packet transmission time for successful reception at the destination. Data packets are transmitted using variable bit rate which is optimized for each link based on the statistics of SINR at destination during past transmitted packets to obtain highest average link data rate. A data packet is successfully received if the SINR at the destination node during the entire packet transmission time is not less than the required SINR for the used data transmission rate. The data packet duration is 1 ms in each scheme and the data packet header and ACK packet overheads are neglected in every scheme. Data packets are generated according to a Poisson process in each source node. The network load is defined as the aggregate bit generation rate in all nodes in the entire network area and is equally distributed among all nodes. Nodes are randomly distributed over the network area and the destination of each node is randomly selected from one of neighboring nodes within distance $d_m$. We evaluate the performance of different schemes using our developed simulations in MATLAB for the following scenarios:
\begin{figure*}
\centering
\subfigure[]{\includegraphics[width=2.3in, trim=.13cm 1.1cm .8cm 1.4cm, clip=true]{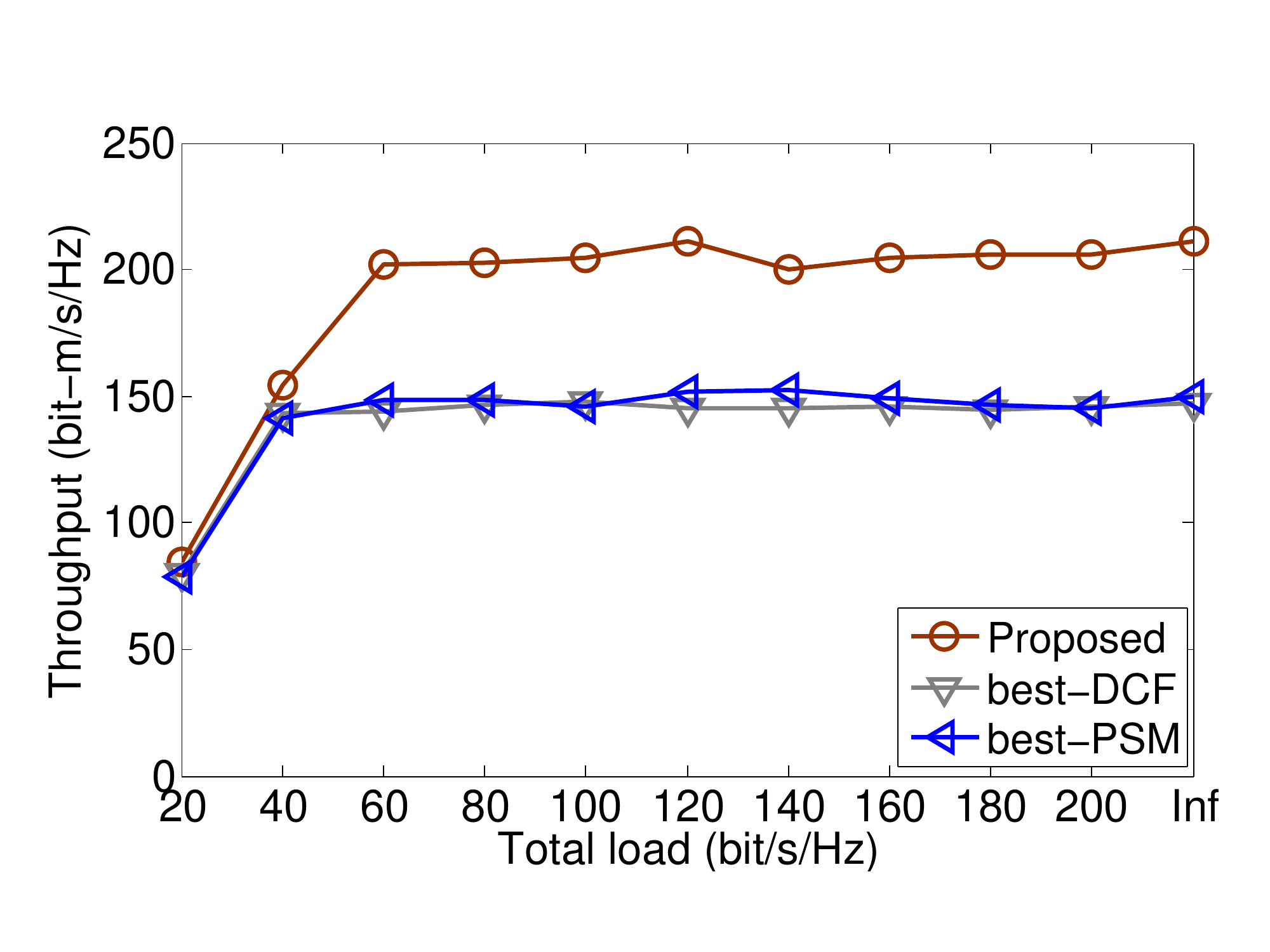}\label{MGTBIT1CH}}
\subfigure[]{\includegraphics[width=2.3in, trim=.2cm 1.4cm 1cm 1.5cm, clip=true]{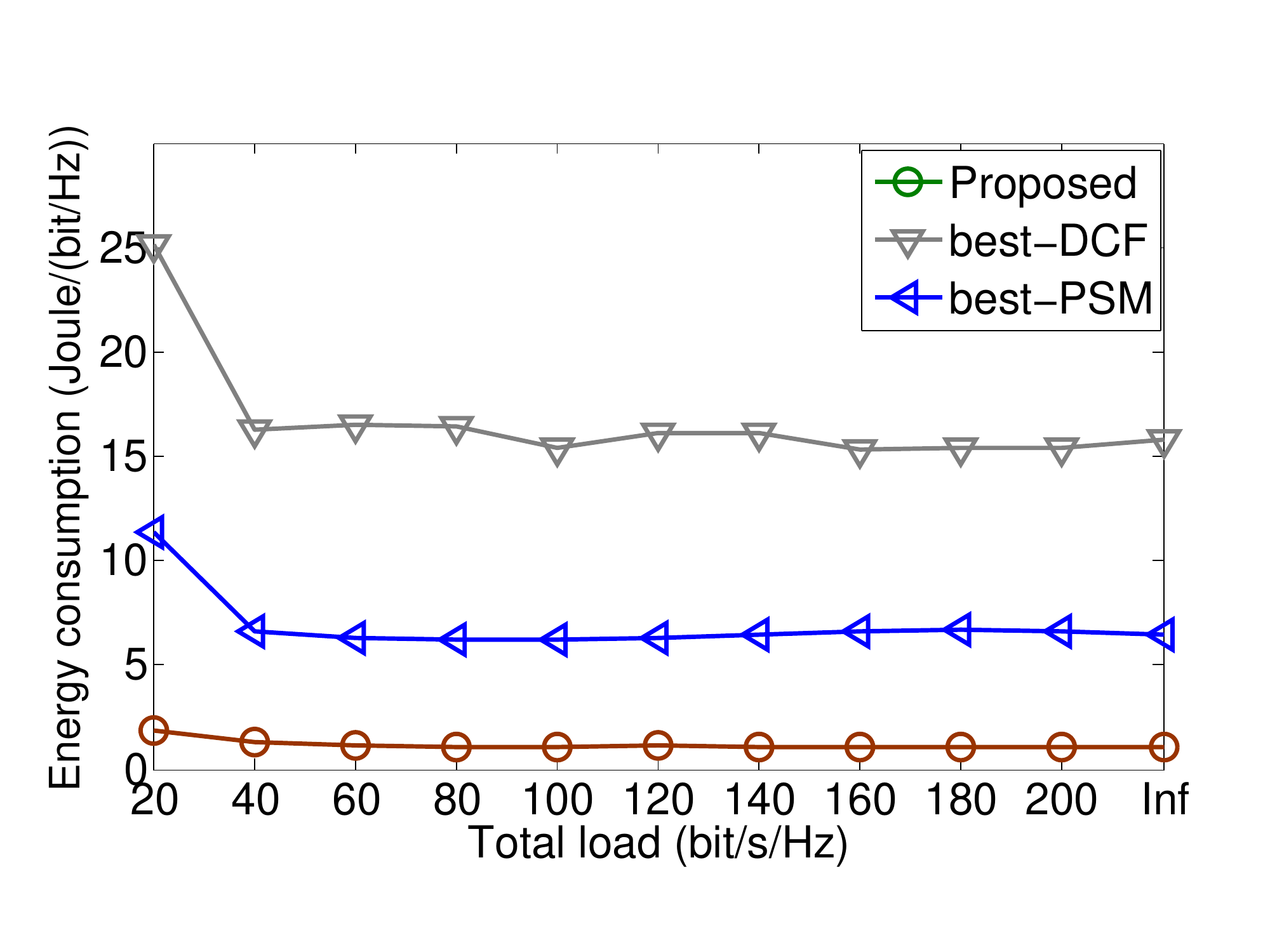}\label{MGTECHperGTBIT0CH} }
\subfigure[]{\includegraphics[width=2.32in, trim=.13cm 1.4cm .5cm 1.5cm, clip=true]{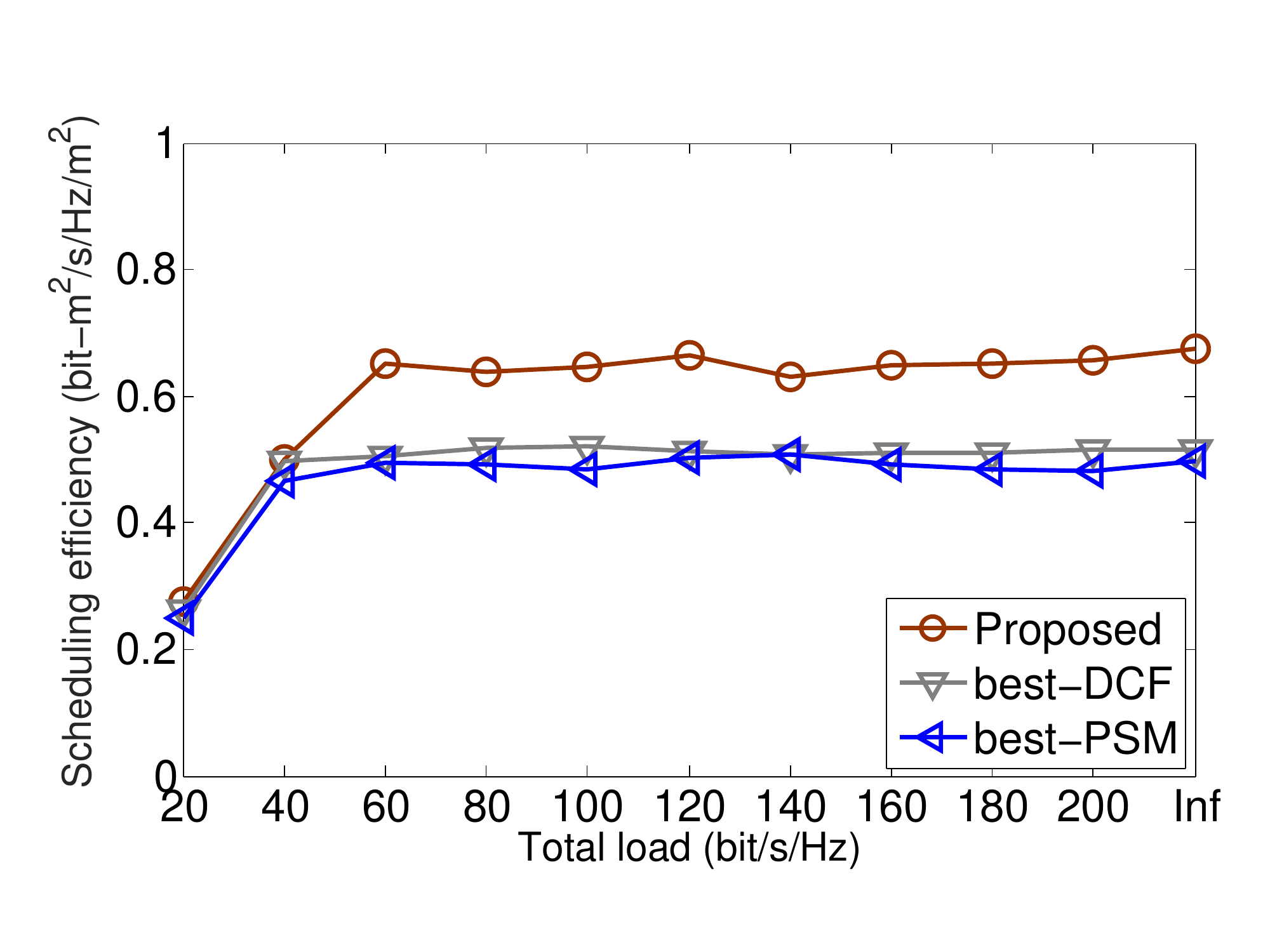}\label{MGTECHperGTBIT0CH}}
\caption{Performance of different schemes in a mobile network scenario as network traffic load varies ($N=400$, $\theta=\infty$).} \label{PMNS}
\end{figure*}
\begin{enumerate}
  \item \begin{it}Static network\end{it} -- Nodes do not move over the simulation time. The simulations are performed for five seconds of the channel time and the performance metrics are averaged over five different random realization of the network;
  \item  \begin{it}Mobile network\end{it} -- Nodes move and network topology varies over the simulation time. Node $i\in\{1,2,...,N\}$ moves with speed $v_i\in [0, 2]$ m/s and along direction $\phi_i\in [-\pi, \pi]$, which are randomly selected for each node with uniform distributions. When the distance between a source and a destination increases to larger than $d_m$, the source node will randomly chooses another destination node. Each node periodically (every one second) reports its current location to the coordinator by transmitting a control packet during contention slots. The simulations are performed for 50 seconds of channel time.
\end{enumerate}
Other simulation parameters are given in Table \ref{C4-Parameters}.

Figures \ref{TBIT1C4vsTEC4-TBIT0}-\ref{GSECHN} show the performance of different schemes in a static network. Figure \ref{TBIT1C4vsTEC4-TBIT0} shows throughput versus energy consumption of the proposed scheme as the energy consumption per bit constraint varies. The energy consumption including only consumed energy during data slots (without considering energy consumed during scheduling slots and contention slots) is also plotted in the figure. According to Figure \ref{TBIT1C4vsTEC4-TBIT0}, as the energy consumption constraints vary from no constrains to the minimum energy consumption per bit constraints in every link, the network throughput is decreased by 38\% and energy consumption is reduced by 18\%, while the energy consumption for data transmissions/receptions only is reduced by 37\%.

Figure \ref{TBIT1-A} shows the throughput using different schemes, as network traffic load and number of nodes change. The data transmission rate of the nodes using different schemes are depicted in figure \ref{TBIT0peNode}.  The proposed scheme provides about 40\% higher throughput than best-DCF and best-PSM. Figure \ref{TBIT1} shows the effectiveness of the strategies used for choosing transmission power and target interference power of the links and for link scheduling in our proposed scheme.

Figure \ref{TBIT0peNode} compares the data transmission rate of the nodes using different schemes. In each scheme, nodes are sorted based on data transmission rate and the horizonal line shows node index. It is observed that the proposed scheme provides better fairness compared to best-DCF and best-PSM, as the link scheduling algorithm in the proposed scheme is to maintain fairness while efficiently choosing concurrent transmissions in each data slot.

The energy consumption using different schemes, as network traffic load and number of nodes change are shown in Figure \ref{TEC-TBIT0-A}. The energy consumption of the proposed scheme is less than 10\% of best-DCF and about 20\% of best-PSM.

Figure \ref{GSECHN} compares the scheduling efficiency using different schemes. The scheduling efficiency of the proposed scheme is about 35\% higher than best-DCF and best-PSM. Indeed, the scheduling efficiency of our proposed scheme is about 70\% of the asymptotic optimal scheduling and transmission power control. The achieved scheduling efficiency is about 78\% in data slots, as 90\% of slots are data slots and the rest are scheduling and contention slots in the proposed scheme.

The performance of different schemes in a mobile scenario is evaluated in Figure \ref{PMNS}. The proposed scheme provides about 30\% higher throughput compared to best-DCF and best-PSM. The energy consumption per transmitted data bit using proposed scheme is less than 20\% of the existing schemes. Also, the scheduling efficiency using the proposed scheme is about 30\% higher than the existing schemes.

\section{Conclusion}\label{Summary-P3}
In this paper, we study joint scheduling and transmission power control for spectrum and energy efficient communication in a wireless ad hoc network. We analyze the asymptotic optimal joint scheduling and transmission power control, and determine the maximum spectrum efficiency, subject to an energy efficiency constraint. Based on the asymptotic analysis, we propose a scheduling and transmission power control scheme to maximize spectrum and energy efficiencies in a practical network. A transmission power level and a target interference power level are determined for each link based on the asymptotic optimal values. Concurrent links are scheduled for transmission such that the actual level of interference at each destination node is close to its target interference level. We present a distributed MAC framework to implement the proposed scheme based on local network information. Simulation results show that the proposed scheme provides about 40\% higher throughput than existing schemes. The energy consumption of the proposed scheme is less than 20\% of existing schemes. Also, the scheduling efficiency of proposed scheme is 70\% of the asymptotic optimal solution, which is about 35\% higher than existing schemes.

\appendix \label{Appendix}

In this section, we discuss solving (\ref{C4-8}) using the method of Lagrangian multipliers. The Lagrangian of (\ref{C4-8}) can be written as
\begin{multline}\label{C4-30}
 L(r_g,\gamma, \mu_1, \mu_2) = \frac{1}{d^2}\times \frac{\log_2\left(1+F(\frac{r_g}{d})\right)}{\frac{3\sqrt{3}}{2}\times\left(\frac{r_g}{d}\right)^2}
 \\ + \mu_1 \times (\displaystyle\frac{2\Gamma_c+g_a \gamma}{\log_2\left(1+F(\frac{r_g}{d})\right)}-\hat{E})
 +\mu_2 \times  (-F(\frac{r_g}{d}) + \eta_{\min}).
\end{multline}
Thus, the KKT conditions can be written as
\begin{multline}\label{C4-31}
\frac{\partial L}{\partial r_g} = \frac{  \frac{ \frac{3\sqrt{3}}{2}\times\left(\frac{r_g}{d}\right)^2 }{\log_e 2 \times (1+F(\frac{r_g}{d})} \times \frac{\partial F(\frac{r_g}{d})}{\partial r_g}  - \frac{3\sqrt{3}r_g}{d^2} \log_2\left(1+F(\frac{r_g}{d})\right) }{\frac{27{r_g}^4}{4d^2}} \qquad
\\ - \mu_1 \times \displaystyle\frac{ (2\Gamma_c+g_a \gamma)\times \log_e 2 \times \frac{\partial F(\frac{r_g}{d})}{\partial r_g} }{\Big(\log_2\left(1+F(\frac{r_g}{d})\right)\Big)^2\times \left(1+F(\frac{r_g}{d})\right)} - \mu_2 \times \frac{\partial F(\frac{r_g}{d})}{\partial r_g}
 \\=0; \qquad \qquad \qquad \qquad \qquad \qquad \qquad \qquad \quad \;\;  \\
 \frac{\partial L}{\partial \gamma} = \mu_1 \times \displaystyle\frac{g_a}{\log_2 \left(1+F(\frac{r_g}{d})\right)} =0 \;; \qquad \qquad \qquad\qquad \\
 \mu_1 \times (\displaystyle\frac{2\Gamma_c+g_a \gamma}{\log_2\left(1+F(\frac{r_g}{d})\right)}-\hat{E}) =0; \qquad \quad \\
 \mu_2 \times  (-F(\frac{r_g}{d}) + \eta_{\min});  \qquad\qquad \qquad \quad \;\\
 \mu_1,\, \mu_2 \geq 0.  \qquad \qquad \qquad \qquad \qquad \qquad \qquad
\end{multline}
The partial derivative $\partial F(r_g/d)/\partial r_g$ in (\ref{C4-31}) can be calculated using (\ref{C4-3}). Finally, the optimal solution can be calculated by examining stationary points in (\ref{C4-31}).

\ifCLASSOPTIONcaptionsoff
 % \newpage
\fi

\bibliographystyle{IEEEtran}
\bibliography{myref}

\end{document}